**Title:** Revising the global biogeography of annual and perennial plants

**Authors:** Tyler Poppenwimer[1,2], Itay Mayrose[1*], Niv DeMalach[2†]

**Affiliations:**
  [1]School of Plant Sciences and Food Security, Tel Aviv University; Tel Aviv, 6997801, Israel.
  [2]Institute of Plant Sciences and Genetics in Agriculture, The Hebrew University of Jerusalem; Rehovot, 7610001, Israel
  *Corresponding author. Email: itaymay@tauex.tau.ac.il
  †Corresponding author. Email: Niv.demalach@mail.huji.ac.il

**Abstract**

There are two main life cycles in plants, annual and perennial[1, 2]. These life cycles are associated with different traits that determine ecosystem function[3, 4]. Although life cycles are textbook examples of plant adaptation to different environments, we lack comprehensive knowledge regarding their global distributional patterns. Here, we assembled an extensive database of plant life cycle assignments of 235,000 plant species coupled with millions of georeferenced datapoints to map the worldwide biogeography thereof. We found that annuals are half as common as initially thought[5, 6, 7, 8], accounting for only 6% of plant species. Our analyses indicate annuals are favored in hot and dry regions. However, a more accurate model shows annual species' prevalence is driven by temperature and precipitation in the driest quarter (rather than yearly means), explaining, for example, why some Mediterranean systems have more annuals than deserts. Furthermore, this pattern remains consistent among different families, indicating convergent evolution. Finally, we demonstrate that increasing climate variability and anthropogenic disturbance increase annual favorability. Considering future climate change, we predict an increase in annual prevalence for 69% of the world's ecoregions by 2060. Overall, our analyses raise concerns for ecosystem services provided by perennials, as ongoing changes are leading to a more annuals-dominated world.

**Introduction**

At the coarsest scale, terrestrial plants can be categorized into two main types of life cycles, annual and perennial[1, 2]. Although crude, this categorization represents the most fundamental characteristic of plant species and illustrates the inherent trade-offs between reproduction, survival, and seedling success[1, 9]. Annual plants reproduce once and complete their life cycle within one growing season, while perennial plants live for many years and, in most cases, reproduce multiple times. The evolutionary trade-offs reflected in these strategies manifest in numerous functional attributes, such as leaf[10] and root traits[11], invasiveness[12, 13], genome characteristics[14], and community stability[15] and therefore have many consequences for ecosystem functioning and services[3, 4]. For example, by allocating more resources belowground, perennials reduce erosion, store organic carbon, and have higher nutrient- and water-use efficiencies[4, 16, 17, 18].

The differences between annual and perennial plants are noticeably reflected in agricultural settings. Despite being a minor part of global biomass[19], annual species are the primary food source of humankind, probably because they allocate more resources to seed output, thereby enhancing agricultural productivity. During the Anthropocene, the global cover of annuals dramatically increased because natural systems, often dominated by perennials, were converted into annual cropland[20, 21]. Annual plants cover ~70% of the croplands and provide ~80% of worldwide food consumption[22]. Moreover, the proportion of annuals increases in many systems because woody perennials have a higher extinction rate[23], while invasive plant species tend to be annuals[12].

The annual life cycle has repeatedly evolved in at least 120 different families, suggesting that it provides a fitness advantage under certain conditions[24]. According to life-history theory, the optimal life cycle is determined by the ratio of seedlings (or seeds) survival to adult survival [25, 26]. The reproductive mode of perennials requires multiple growing seasons[1] compared to annuals which require only one growing season. Therefore, any external condition that decreases the ability of plants to survive between growing seasons necessarily reduces the reproductive fitness of perennial species[25, 26]. However, because annual species could survive such conditions as seeds rather than adults, their reproductive fitness may not be impacted[1]. Thus, any condition that skews the survivorship ratio in favor of seeds should increase the favorability of annuals.

Consequently, annuals should be favored when adult mortality is high and seed persistence and seedling survival are relatively high.

Numerous studies have discussed plant life cycles as primary examples of adaptation to different climatic conditions and provide estimates for their prevalence in various regions[5, 6, 7, 8]. However, the data provided in many of these studies, which penetrated many current ecological textbooks[7, 8], are problematic in several aspects. First, the current estimate for the global proportion of annual species (13%) is based on a century-old sample of merely 400 species[2], representing 0.1% of accepted plant species[27]. Second, current biome level estimates are based on a single location and extrapolated to represent the entire biome. For example, the desert biome is assumed to contain 42% annual species[6, 7] but is based only on data from Death Valley in California[2]. Third, estimates are inconsistent and difficult to compare due to ambiguous biome definitions. For example, an alternative estimate for the desert biome suggests that 73% of plant species are annuals[5, 8]. Lastly, each biome incorporates a wide range of conditions, e.g., the mean temperature in the desert biome ranges from 30° to -10°C, corresponding to hot and cold deserts. Thus, this definition aggregates regions that differ markedly in their environmental conditions, likely affecting the prevalence of the different life cycles.

As central as life cycles are to plant ecology and evolutionary research, it is remarkable that we still have no precise estimate for the worldwide prevalence of life cycles and their environmental drivers. Yet, such an assessment is essential in times of climate and land-use changes[20, 28], which are expected to dramatically alter patterns of plant biogeography with many consequences for ecosystem processes and services[29, 30, 31, 32]. Here, we present a comprehensive assemblage of plant life cycle data encompassing over 235,000 plant species. We cross this database with millions of georeferenced data points to produce the first worldwide map of plant life-cycle distribution. This extensive plant growth-form database contains life cycle data for 67% of all vascular plant species and georeferenced data for 51%. These data allow us to evaluate the underlying drivers of plant life cycle strategies that were never tested on a global scale. We tested three key hypotheses, predicting that annuals are favored under: (1) increasing temperature and decreasing precipitation[24, 33, 34, 35], (2) high year-to-year variability in climatic conditions[35, 36, 37], and (3) increasing human footprint (anthropogenic disturbance[36, 38, 39, 40]). All these hypotheses are based on the life-history theory that predicts annual species to be favored with increasing adult mortality (relative to seedling mortality)[25, 26]. In other words, the relative abundance of annuals will be higher in regions with hot-dry climates, high interannual variability, and disturbance because they decrease adult survival. Finally, with a more accurate understanding of the global drivers, we provide an initial assessment regarding the impact of future conditions on plant life cycle distribution.

**Results and Discussion**

This compilation of life cycles revealed dramatic differences in the relative prevalence of annual and perennial plant species compared to existing estimates. Annual species comprise 6% of all species and 15% of herbaceous species (i.e., omitting all woody species, which are all perennial). Moreover, only 5.5% of ecoregions exhibit an annual herbs proportion of 50% or more (Fig. 1).

Below, we focus on the proportion of annual species among herbaceous species (rather than among all species), which provides a better 'resolution' in regions with a high proportion of woody species. Nonetheless, similar results were obtained when we analyzed the proportion of annuals among all species (Supplement Note 2 & Extended Data Fig. 1).

The variation in the annual-herb frequencies across biomes supports the first hypothesis that annuals are favored with increasing temperature and lower precipitation (Fig. 2). Still, the differences among biomes (following Whittaker's[41] approach) were not substantial (Fig 2A, Table 1). The proportion of annual herbs ranges from 13% to 25%, suggesting that the role of climate is underestimated in this coarse spatial scale. The large variability within each biome is revealed when examining the proportion of annuals at the ecoregion resolution (Fig. 2B). For example, not all desert-biome ecoregions have a high proportion of annual herb species, with cooler deserts exhibiting much lower proportions than hot ones. The same trend is repeated among other biomes, as ecoregions with lower precipitation and hotter temperatures (i.e., located in the lower-left coordinate of their biome in Fig. 2B) possess a greater proportion of annuals.

This pattern was corroborated using a linear regression model that fitted the proportion of annuals as a function of *mean yearly temperature* and *total yearly precipitation*. These two climatic variables accounted for nearly half of the variance of the worldwide distribution of plant life cycle strategies ($P < 10^{-15}$, $D.F. = 679$, $R^2 = 0.48$). As mean yearly temperature increases and total precipitation decreases, the proportion of annuals increases (Fig. 2C). These results are robust to spatial autocorrelation, with only negligible differences in the parameter estimates and correspondingly low p-values (Supplement Note 3) and to alternative statistical methods such as Poisson regression (Supplement Note 4, Extended Data Fig 2).

Although yearly temperature and precipitation provide a good description of annual herb proportions across the globe, it does not account for temporal variation in climate throughout the year. We, therefore, fitted a suite of two-variable regression models. Each model consisted of one *quarterly* temperature variable and one quarterly precipitation variable. The best-fit model (hereafter the quarterly model) incorporated the *mean temperature of the warmest quarter* and the log-transformed *precipitation of the warmest quarter* and accounted for 55% of the observed variance ($P < 10^{-15}$, $D.F. = 679$). According to this model, annual herbs proportion increases with increasing temperature and decreasing precipitation of the warmest quarter (Fig. 3). Furthermore, this model had a substantially better fit than the model based on the *mean yearly temperature* and *total yearly precipitation* outperforming it in terms of explained variance (0.55 vs. 0.48) and information theory criteria ($\Delta AICc = 92.4$). These results provide a more nuanced understanding of the first hypothesis, demonstrating that hot and dry conditions impact the prevalence of annuals, particularly in the driest season.

The quarterly model can distinguish between ecoregions with similar yearly climate patterns yet a different proportion of annuals. For example, the Eastern Mediterranean (e.g., Tel Aviv) and Chihuahuan desert (southwestern United States and northern Mexico) ecoregions have identical mean yearly temperatures (17.6° C) and relatively similar amounts of yearly precipitation (527mm and 330mm), yet maintain different annual herb proportions (51% and 36%). Given their similar mean yearly climate, the yearly temperature and precipitation model predicts similar annual herb proportions for these two ecoregions (32% and 34%, respectively). However, the Tel Aviv ecoregion receives substantially less precipitation (6mm) than the Chihuahuan desert (157mm) during the hottest quarter. As such, the quarterly model differentiates the two ecoregions, producing substantially better predictions (annual herbs proportion of 52% in Tel Aviv and 29% in the Chihuahuan desert). Consequently, the coinciding of high-temperature and low-precipitation periods increases the favorability of annuals more than simply yearly means.

We conducted two analyses to account for potential biases of the revealed trends due to phylogenetic dependence. First, using the quarterly model, we conducted a separate analysis for the four most annual-rich families (Asteraceae, Brassicaceae, Fabaceae, and Poaceae).

Qualitatively similar relationships between climate and annual proportion were found in all families (Fig 4), providing evidence for convergent evolution of annual life cycles in hot and dry conditions. Next, we tested the life cycle and climate relationship using phylogenetic Generalized Least Squares (pGLS). We found that the median temperature of the warmest quarter for annuals is 3°C higher, and the median precipitation of the warmest quarter is 35% lower (Supplement Note 5). These results support the hypothesis that climate conditions during the driest period play a significant role in driving the prevalence of annuals.

We tested the second hypothesis that increased year-to-year climatic variability favors annuals prevalence by focusing on interannual variability in total precipitation (in terms of the coefficient of variation) and mean temperature (in terms of standard deviation). Using a bivariate regression, we found that increasing precipitation variability is associated with a higher proportion of annual species ($P < 10^{-15}$, $D.F. = 679$, $R^2 = 0.24$). Likewise, we found that increasing temperature variability also increases the favorability of annuals, though its effect is much weaker ($P = 0.0003$, $D.F. = 679$, $R^2 = 0.02$). Furthermore, incorporating precipitation and temperature inter-annual variability into the quarterly model improved model fit (from $R^2 = 0.55$ to $R^2 = 0.61$) and overall performance ($\Delta AICc = 51$).

We examined our third hypothesis that increased human footprint (anthropogenic disturbance) should increase the proportion of annuals. In a bivariate regression, we found a positive effect of the human footprint on the proportion of annuals ($P < 10^{-7}$, $D.F. = 680$, $R^2 = 0.04$). However, despite the mild individual explanatory power of human footprint, adding it to the previous model with the four climatic variables further improved the model's explanatory power ($\Delta AICc = 15$, change in $R^2$ from 0.61 to 0.63).

Finally, we built a back-of-the-envelope projection of the expected prevalence of annuals in 2060 based on predicted changes in mean temperature and precipitation at the warmest quarter[42] (Extended Data Fig 3). Under the simplifying assumptions that the prevalence of annuals in the future will follow the same climatic patterns without adaptation or time-lag, our model suggests that ~69% of ecoregions will experience an increase in the proportion of annuals.

## Conclusions

This study provides an extensive update to the worldwide biogeography of plant life cycles and demonstrates major differences from previous estimates. At the global level, our analyses indicate that annual species are half as common as previously thought[5, 6, 7, 8]. Similarly, our estimates at the biome-level vary from earlier estimates changing some by as much as 3-5 fold (Table 1 and Extended Data Table 1). Additionally, these revised estimates display a more limited difference between the biome with the highest and lowest annual proportion, reducing the difference from 60% to a more restricted 12%.

Overall, our analyses provide general support for our three hypotheses regarding the conditions under which annual proportions will increase. First, we find that the proportion of annuals increases under hotter and drier conditions, and this result is robust to spatial autocorrelation and phylogenetic relatedness. However, yearly means provide an insufficient explanation for some observed patterns. After exploring alternative climate patterns, we determined that a long dry summer is a principal factor governing the occurrence of annual-rich regions, demonstrating that the temporal distribution of hot and dry periods is more important than having an arid climate *per se*.

Second, our results suggest that annuals are more prevalent under increasing climate unpredictability. and interannual temperature variability increase, the proportion of annuals also increases. However, the correlation between temperature variability and annual proportion is weaker than precipitation variability, indicating that irregular precipitation patterns have a larger impact.

Thirdly, our findings demonstrate that as human-mediated disturbance increases, the favorability of annual plants also increases. Furthermore, we found that a substantial portion of the effect of human disturbance is independent of climatic patterns. Although there is extensive evidence that human disturbance enhances the abundance of annuals in local communities[38], our study is the first, to our knowledge, to provide evidence that human disturbance favors annuals on a biogeographical scale.

Finally, our future projection model predicts that by 2060, we will experience an increase in the prevalence of annuals. However, we caution that our back-of-the-envelope prediction is based on the simplistic assumption that biogeographic patterns instantly track climate changes (i.e., it does not account for time lags in species response to a changing climate). Still, our prediction is also conservative in the sense that it does not account for the predicted increase in year-to-year climatic variability[42] as well as human footprint. With the human population predicted to reach 10 billion by 2060, anthropogenic activities are expected to play an increasing role in shaping patterns of plant biogeography. Consequently, we expect a world with more annual-dominated ecoregions.

## References

1. Friedman, J. The evolution of annual and perennial plant life histories: ecological correlates and genetic mechanisms. *Annu. Rev. Ecol. Evol. Syst.* **51,** 461–481 (2020).

2. Raunkiær, C. *Über das biologische normalspektrum*. (Andr. Fred. Host & Son, Kgl. Hof-Boghandel, 1918).

3. Glover, J. D., *et al*. Increased food and ecosystem. *Science.* **328,** 1638–1640 (2010).

4. Vico, G., Manzoni, S., Nkurunziza, L., Murphy, K., Weih, M. Trade-offs between seed output and life span – a quantitative comparison of traits between annual and perennial congeneric species. *New Phytol.* **209,** 104–114 (2016).

5. Whittaker, R. H., Lovely, R. A. *Communities and ecosystems* (Macmillan, New York, 1975).

6. Crawley, M. J. *Plant ecology* (Blackwell Science, Oxford, 1997).

7. Begon, M., Townsend, C. R. *Ecology: from individuals to ecosystems* (John Wiley & Sons, Hoboken, 2021).

8. Gurevitch, J., Scheiner, S. M., Fox, G. A. *The ecology of plants* (Oxford University Press, New York, 2021).

9. Salguero-Gómez, R., *et al*. Fast-slow continuum and reproductive strategies structure plant life-history variation worldwide. *Proc. Natl. Acad. Sci. U. S. A.* **113,** 230–235 (2016).

10. Garnier, E., Vancaeyzeele, S. Carbon and nitrogen content of congeneric annual and perennial grass species: relationships with growth. *Plant. Cell Environ.* **17,** 399–407 (1994).

11. Roumet, C., Urcelay, C., Díaz, S. Suites of root traits differ between annual and perennial species growing in the field. *New Phytol.* **170,** 357–368 (2006).

12. Funk, J. L., Standish, R. J., Stock, W. D., Valladares, F. Plant functional traits of dominant native and invasive species in mediterranean-climate ecosystems. *Ecology*. **97,** 75–83 (2016).

13. Murray, B. R., Thrall, P. H., Gill, A. M., Nicotra, A. B. How plant life-history and ecological traits relate to species rarity and commonness at varying spatial scales. *Austral Ecol.* **27,** 291–310 (2002).

14. Rice, A., *et al*. The global biogeography of polyploid plants. *Nat. Ecol. Evol.* **3,** 265–273 (2019).

15. Grman, E., Lau, J. A., Schoolmaster, D. R., Gross, K. L. Mechanisms contributing to stability in ecosystem function depend on the environmental context. *Ecol. Lett.* **13**, 1400–1410 (2010).

16. Glover, J. D., Reganold, J. P., Cox, C. M. Plant perennials to save Africa's soils. *Nature*. **489,** 359–361 (2012).

17. Kreitzman, M., Toensmeier, E., Chan, K. M. A., Smukler, S., Ramankutty, N. Perennial staple crops: yields, distribution, and nutrition in the global food system. *Front. Sustain. Food Syst.* **4,** 1–21 (2020).

18. Ledo, A., *et al*. Changes in soil organic carbon under perennial crops. *Glob. Chang. Biol.* **26,** 4158–4168 (2020).

19. Bar-On, Y. M., Phillips, R., Milo, R. The biomass distribution on Earth. *Proc. Natl. Acad. Sci. U. S. A.* **115,** 6506–6511 (2018).

20. Foley, J. A., *et al*. Global consequences of land use. *Science.* **309,** 570–574 (2005).

21. Erb, K. H., *et al*. Unexpectedly large impact of forest management and grazing on global vegetation biomass. *Nature*. **553,** 73–76 (2018).

22. Pimentel, D. *et al*. Annual vs. perennial grain production. *Agric. Ecosyst. Environ.* **161,** 1–9 (2012).

23. Humphreys, A. M., Govaerts, R., Ficinski, S. Z., Nic Lughadha, E., Vorontsova, M. S. Global dataset shows geography and life form predict modern plant extinction and rediscovery. *Nat. Ecol. Evol.* **3,** 1043–1047 (2019).

24. Friedman, J., Rubin, M. J. All in good time: Understanding annual and perennial strategies in plants. *Am. J. Bot.* **102,** 497–499 (2015).

25. Cole, L. C. The population consequences of life history phenomena. *Q. Rev. Biol.* **29,** 103–137 (1954).

26. Charnov, E. L., Schaffer, W. M. Life-history consequences of natural selection: Cole's result revisited. *Am. Nat*. **107,** 791–793 (1973).

27. WFO. World flora online. Published on the Internet at http://www.worldfloraonline.org. (2023).

28. Díaz, S., *et al*. Pervasive human-driven decline of life on Earth points to the need for transformative change. *Science.* **366,** (2019), doi:10.1126/science.aax3100.

29. Grimm, N. B., *et al*. The impacts of climate change on ecosystem structure and function. *Front. Ecol. Environ.* **11,** 474–482 (2013).

30. Hooper, D. U., *et al*. Effects of biodiversity on ecosystem functioning: a consensus of current knowledge. *Ecol. Monogr.* **75,** 3–35 (2005).

31. Chapin III, F. S., *et al*. Consequences of changing biodiversity. *Nature*. **405,** 234–242

(2000).

32. Weiskopf, S. R., *et al*. Climate change effects on biodiversity, ecosystems, ecosystem services, and natural resource management in the United States. *Sci. Total Environ.* **733,** (2020), doi:10.1016/j.scitotenv.2020.137782.

33. Datson, P.M., Murray, B.G. & Steiner, K.E. Climate and the evolution of annual/perennial life-histories in *Nemesia* (Scrophulariaceae). *Plant Syst Evol*. **270,** 39–57 (2008).

34. Evans, M. E. K., Hearn, D. J., Hahn, W. J., Spangle, J. M., Venable, D. L. Climate and life-history evolution in evening primroses (Oenothera, Onagraceae): a phylogenetic comparative analysis. *Evolution.* **59,** 1914–1927 (2005).

35. Zeineddine, M., Jansen, V. A. A. To Age, To die: parity, evolutionary tracking and Cole's paradox. *Evolution*. **63,** 1498–1507 (2009).

36. Cruz-Mazo, G., Buide, M. L., Samuel, R., Narbona, E. Molecular phylogeny of Scorzoneroides (Asteraceae): evolution of heterocarpy and annual habit in unpredictable environments. *Molecular phylogenetics and evolution*. **53,** 835-847. (2009).

37. Murphy, G. I. Pattern in life history and the environment. *The American Naturalist*. **102,** 391-403. (1968).

38. Díaz, S., *et al*. Plant trait responses to grazing - A global synthesis. *Glob. Chang. Biol.* **13,** 313–341 (2007).

39. Herben, T., Klimešová, J., Chytrý, M. Effects of disturbance frequency and severity on plant traits: an assessment across a temperate flora. *Funct. Ecol.* **32,** 799–808 (2018).

40. Pianka, E. R. On r-and K-selection. *The American Naturalist*. **104,** 592-597. (1970).

41. Whittaker, R. H. Communities and ecosystems. *Communities and Ecosystems.* (1970).

42. Salinger, M.J. Climate variability and change: past, present and future — an overview in *Increasing climate variability and change* (eds. Salinger, J., Sivakumar, M., Motha, R.P.) 9-29 (Springer, 2005). 313–341 (2007).

# Figures and Tables

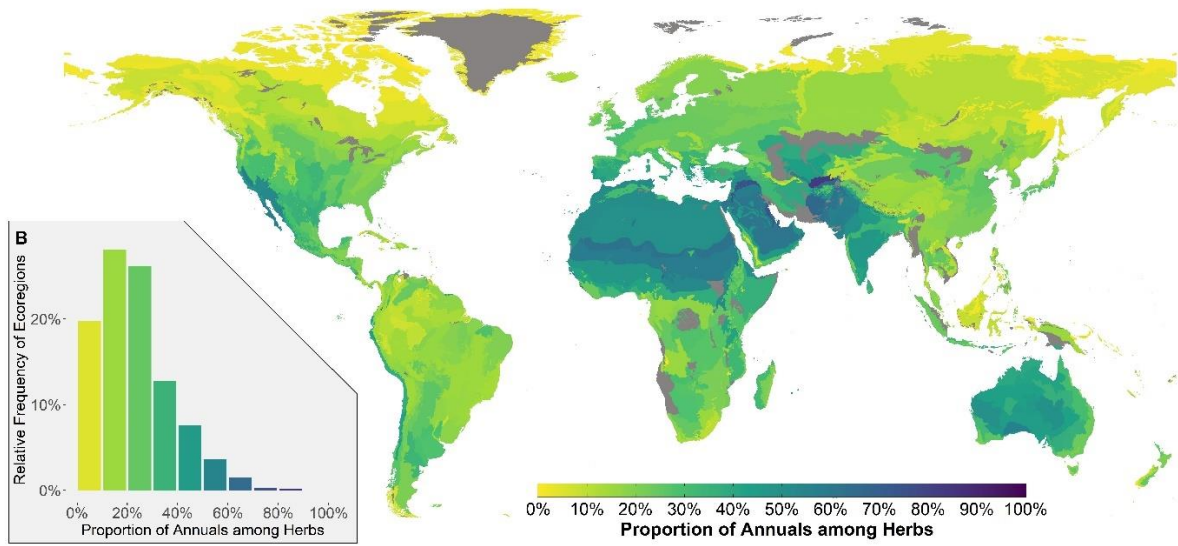

**Figure 1 | The worldwide biogeography of annuals prevalence. A**, A map of the proportion of annual species (among herbaceous species) in each ecoregion. **B**, The distribution of annual proportions among ecoregions. Ecoregions with insufficient data (see Methods) are colored grey resulting in 682 colored ecoregions.

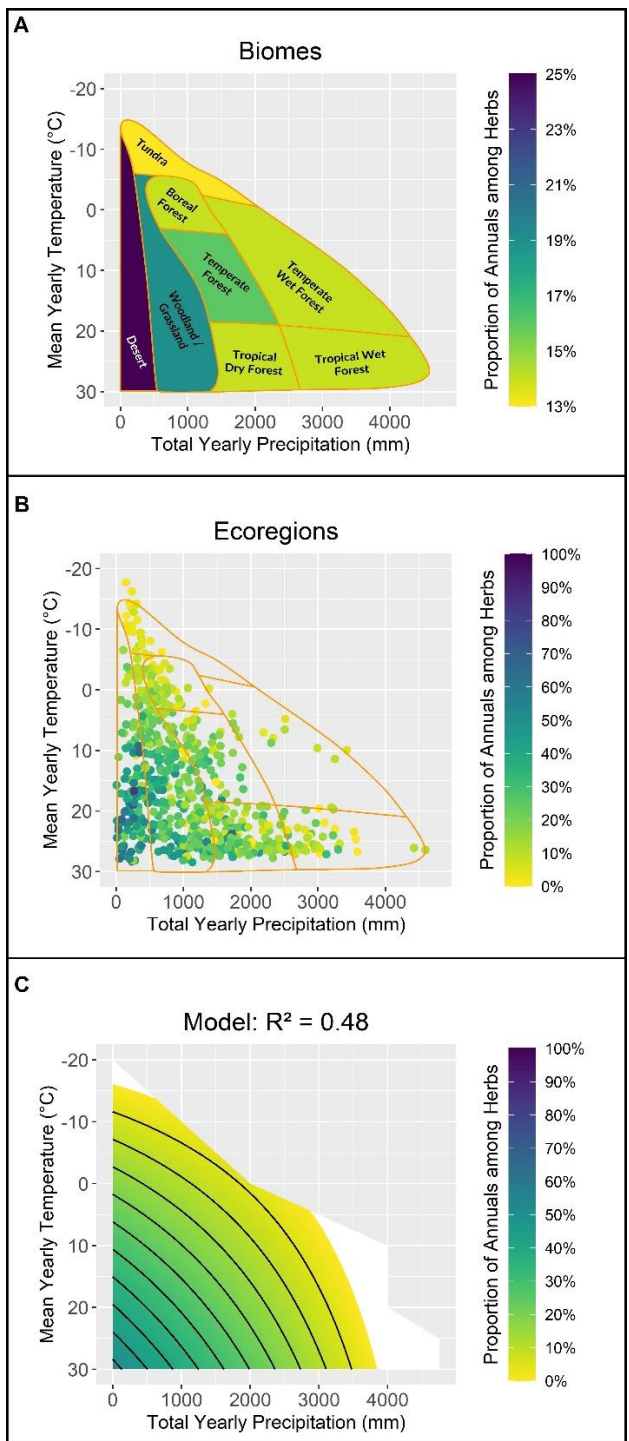

**Figure 2 | The effects of total yearly precipitation and mean yearly temperature on the proportion of annuals (among herbaceous species). A**, The proportion of annuals in each of Whittaker's biomes[41]. **B,** The proportion of annuals in each ecoregion (the outline of Whittaker's biomes is marked by orange lines). **C,** Predictions of a linear regression model of the proportion of annuals as a function of the mean yearly precipitation and temperature (contour lines every 5%). Note that the scale is different for panel **A**. N=682.

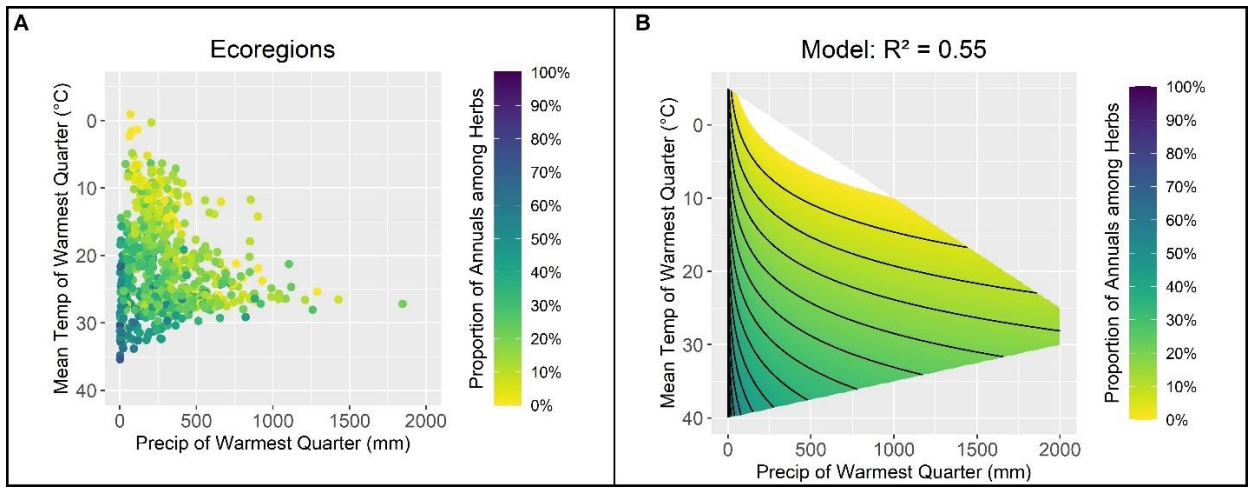

**Figure 3 | The effects of the precipitation and mean temperature of the warmest quarter on the proportion of annuals (among herbaceous species). A**, The proportion of annuals in each ecoregion. **B**, The predictions of a linear regression model of annual proportion as a function of precipitation and mean temperature of the warmest quarter with contour lines every 5%. N=682.

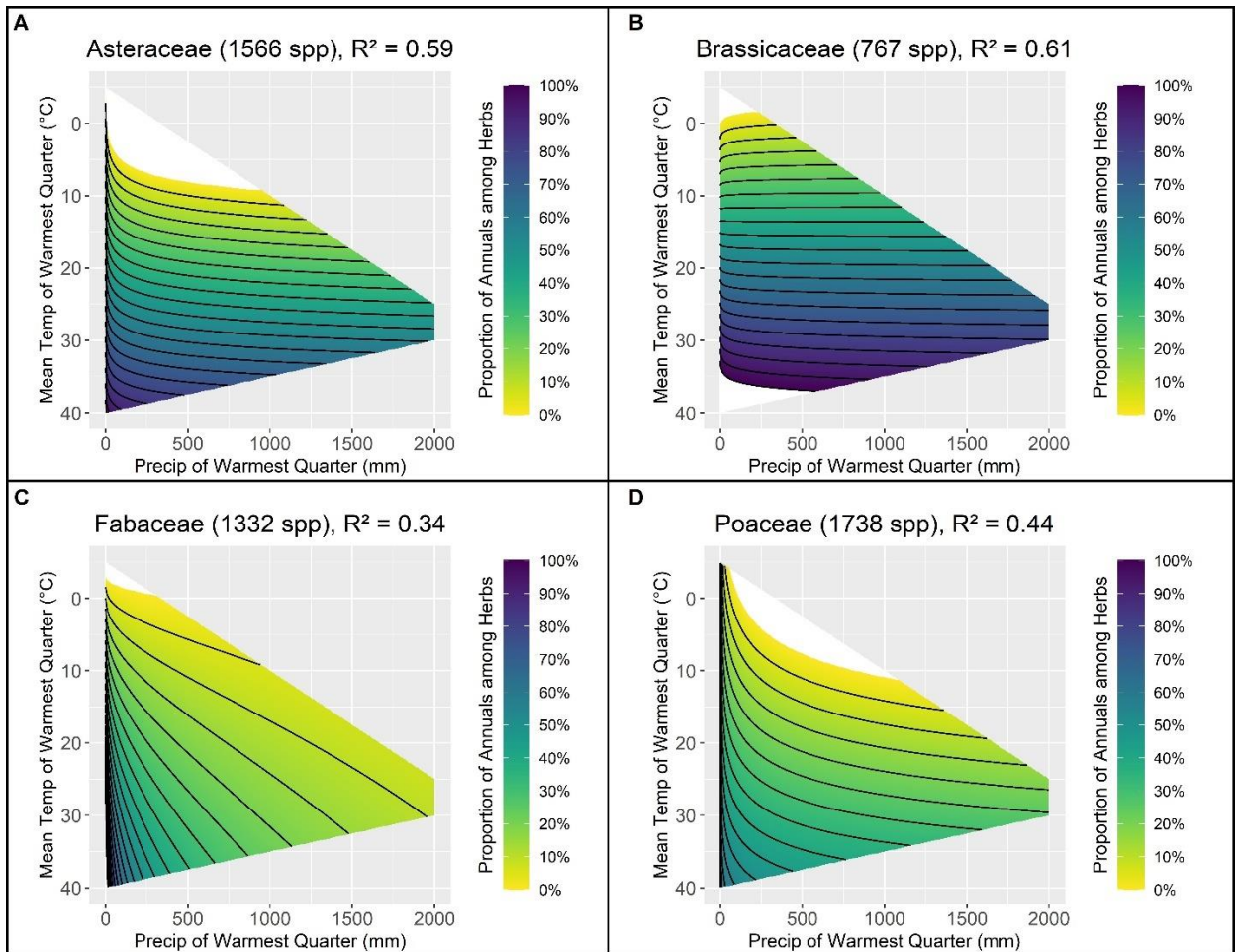

**Figure 4 | The effects of the precipitation and mean temperature of the warmest quarter on the proportion of annuals (among herbaceous species) in the four most annual-rich families (predictions of the linear regression model). A**, Asteraceae, N = 465. **B**, Brassicaceae N = 262. **C**, Fabaceae N = 382. **D**, Poaceae N = 513. Contour lines are drawn every 5% for all figures.

**Table 1 | A comparison of previous estimates, obtained from[7], for the proportion of annuals among all species and among herbaceous species to our revised estimates.** Greyed cells have no initial biome estimate. Alternative previous estimates are available in Extended Data Table 2. Note, the biome nomenclature used for the previous estimates differs from ours and so the location of the original study was used to determine the corresponding biome. Additional information can be found in Extended Data Table 3.

| Region | Annuals among all species | | Annuals among herbaceous species | |
| --- | --- | --- | --- | --- |
| | Previous Estimate | **Revised Estimate** | Previous Estimate | **Revised Estimate** |
| **Global** | **13%** | **6%** | **28%** | **13%** |
| Desert | 42% | **14%** | 63% | **25%** |
| Tundra | 2% | **11%** | 3% | **13%** |
| Woodland / Grassland | 39% | **9%** | 53% | **19%** |
| Boreal forest | | **11%** | | **14%** |
| Tropical Dry Forest | | **3%** | | **14%** |
| Tropical Wet Forest | 16% | **3%** | 44% | **14%** |
| Temperate Forest | 18% | **9%** | 20% | **16%** |
| Temperate Wet Forest | | **7%** | | **14%** |

## Methods
### Life Cycle database development

We built an extensive life-cycle database by aggregating all types of vascular plant data from 11 disparate plant trait databases[14, 43, 44, 45, 46, 47, 48, 49, 50, 51, 52] (see Supplement Note 6 for access dates). This raw database contained ~6.4 million entries and 400,000 unique names. All unique names were resolved using the R package *WorldFloraOnline* v1.7[53] (WFO) to ensure a uniform naming scheme and to exclude unrecognized species. Resolved names were filtered by match distance and WFO acceptance (see Supplement Note 7 for a full description).

Following name resolution, each entry consisted of a single species name and its associated trait term (e.g., annual, forb/herb, tree, 10 years, Shrub/Herb, aquatic, TREE, epiphyte, etc.). All unique trait terms were manually assessed to extract data relevant to a plant's life cycle (annual/perennial) and growth form (woody/herbaceous) when available. Those that did not provide relevant information (e.g., Terrestrial_Trailing_Plant, 2.4, NO, b H, etc.) or provided conflicting information for the same entry were excluded (e.g., Shrub/Herb, Tree/Terrestrial Herb, etc.). After term interpretation, there were 5.6 million entries and 262,000 unique *species* remaining.

Life cycle consensus among each species' data was achieved by comparing all life cycle and biomass composition entries for that species. Only those species with a unanimous term agreement and without conflicting life cycle and biomass composition consensuses were considered. Crop species were excluded as occurrence data may not represent natural habitats. A list of crop species was obtained from[14]. This process produced a database of 235,979 species with life cycle information. Our database contains 67% of all WFO-accepted plant species names and represents the largest plant life cycle database developed to date.

### Matching life cycle data with species observations

Species observation data were based on occurrence data from the Geographic Biodiversity Information Facility[54] (GBIF). All observation data points within the Plantae kingdom (~355 million) were downloaded (September 14th, 2021) and processed locally. We filter unreliable data points following the recommendation provided in the vignette of the R package *CoordinateCleaner* v2.0-18[55]. The following steps were used to filter unreliable data points:

1) Data points without coordinates were excluded.

2) The R package *CoordinateCleaner* v2.0-18[55] was used to discard data points with wrong locations and problematic temporal metadata (see Supplement Note 8 for a full description of this process).
3) Data points were removed if the recorded 'coordinate uncertainty' was greater than 100km.
4) Data points whose 'Basis of records' was literature or living specimen were discarded (these generally refer to the location of museum or herbaria collections)
5) Data points whose record date was during or before 1945 were excluded as it has been suggested these may be less reliable.
6) Data points that were not labeled as species.

Once cleaned, all remaining unique names were resolved using the WFO package[53], and the same criteria as in the life form database were applied. Once the names were resolved, the species in the cleaned GBIF database and the assembled life form database were matched. Of the 235,979 species in our assembled lifeform database, 182,848 species were found within the cleaned GBIF data.

To mitigate sampling bias and inexact coordinates, species observation data were mapped into larger geographical regions defined by specific environmental and ecological conditions[14]. To this end, each georeferenced data point was assigned to one of 827 ecoregions as defined by the World Wildlife Fund[56] (WWF). This process was accomplished using the R packages *raster* v3.4-13[57] and *rgdal* v1.5-27[58]. Following the procedures used by[14], species were only considered "present" in a geographic region if there were five or more observations to ensure the species had a sufficient established population. Similarly, to ensure all regions contained sufficient data for analysis, each region was only considered if ten or more species were present.

This procedure produced sufficient data for 723 ecoregions when examining annual species among all species and 682 ecoregions for annual species among only herbaceous species.

We additionally analyzed the data based on a grid system (using 100km × 100km cells) and found similar results to our main ecoregion-based analyses. Further details of these analyses are provided in (Supplement Note 9 and Extended Data Fig 4).

**Predictors of annual proportion**

We examined the relationships between various climatic and anthropogenic features and the distribution of plant life cycle strategies. To this end, we determined the frequency of plant life form strategies by considering the number of species with a given trait out of the total number of species with life form data in each region (e.g., annual species out of all species with life cycle data). Each region was subsequently assigned a suite of climatic and anthropogenic features. Unless otherwise indicated, all features were determined by taking the median value across all pixels in a region.

We downloaded bioclimate features from the WorldClim Global Climate Data[59], which were developed from climate data during 1970 – 2000, at ten arc-minutes resolution. All 19 BIOCLIM variables representing each region's major temperature and precipitation characteristics were extracted using the R package *raster* v3.4-13[57].

As a measure of climate unpredictability, we measured interannual precipitation variation (IPV) and interannual temperature variation (ITV). The IPV metric was obtained by extracting the coefficient of variation from ecoregion precipitation. The ITV metric was obtained by extracting the standard deviation from the ecoregion temperature using the R package *raster* v3.4-13[57]. We aggregated all available monthly precipitation/temperature data layers from the WorldClim Global Climate Data[59] at ten arc-minutes resolution (1961 – 2018) to determine the total yearly precipitation and mean temperature for each pixel in each year. The mean and coefficient of variation of the total yearly precipitation across all 58 years for each pixel were then used to determine IPV values. Similarly, the mean and standard deviation of the yearly mean temperature across all years for each pixel were then used to determine ITV values (for temperature, the coefficient of variation is inappropriate because it inflates values near zero temperature). Unlike other available bioclimatic features, such as BIOCLIM 3, which determine variability within a year, our estimates measure variability between years.

We used the Human Footprint data layer[60] as a proxy for anthropogenic disturbance, representing the human population's total ecological footprints. This layer incorporates eight variables: built-up environments, population density, electric power infrastructure, crop lands, pasture lands, roads, railways, and navigable waterways. Together, these features evaluate the amount of land or sea necessary to support human activity's consumption habits. Human Footprint values were extracted for each region using the R package *raster* v3.4-13[57].

**Biome Estimates**

To obtain biome estimates of annual and annual herb frequencies, all ecoregions with sufficient data were individually plotted in the total yearly precipitation and mean yearly temperature space of the Whittaker biome overlay outline (adapted from[41]) overlaid (see Fig 2B for reference). We determined each ecoregion's biome based on its location within this space.

For those ecoregions whose biome designation was difficult to assess, their points were enlarged until one biome had a plurality of the circle's area. For those ecoregions outside Whittaker's biome space, their biome designation was determined by the closest biome. Once the biome designation of all ecoregions was determined, the species presence data for all ecoregions within a given biome were aggregated. The same process used to determine the presence and absence of species in an ecoregion was used to determine the presence and absence of species in the biome. Of note, a biome could have more species than the combined ecoregions within said biome because some species may have five or more observations within the biome, but not within any of the individual ecoregions.

Whittaker's defined biomes were chosen to simplify comparisons to previous estimates (the terminology used in textbooks best matched those of Whittaker's definitions) and for its simplicity of using only temperature and precipitation.

**Comparing Previous Biome Estimates**

The classification approach for biomes used in previous estimates of annual proportions was not explicitly defined, making a direct comparison with our set of biomes difficult. However, we traced the origins of each estimate and determined the original study's locations. These locations were then matched with the WWF ecoregions, and the corresponding biome was determined as discussed above. This procedure allowed a direct comparison between previous estimates and our revised estimates. Additionally, previous studies did not explicitly provide estimates for the proportion of annuals among herbaceous species. Therefore, for comparison purposes, previous annual herbs proportion estimates were calculated based on the biome-level life form classification estimates from each study. See Extended Data Tables 1–4 for original study location matchings and annual herbs proportions calculations.

**Statistical analyses**
<u>Temperature and Precipitation</u>
To assess support for our first hypothesis, we linearly regressed annual and annual herb frequency against mean yearly temperature, total yearly precipitation, and their interaction. Subsequently, we compared models based on two climatic variables, using one quarterly temperature bioclimatic variable and one quarterly precipitation variable, to identify a potentially better model. We used four temperature bioclimatic features (BIO8, BIO9, BIO10, BIO11) and four precipitation features (BIO16, BIO17, BIO18, BIO19). Month-specific bioclimatic features were omitted because they are highly correlated with quarter-specific features. Preliminary analysis suggested that log transformations of precipitation bioclimatic features often increased explanatory power, and therefore they were also included in the exhaustive search.

Altogether, this grouping scheme produced 32 different 2-feature linear regression models (four temperature and eight precipitation features) with an additional two linear regression models using mean yearly temperature and total yearly precipitation and the log transformation of total yearly precipitation. Model comparison was achieved using AIC values obtained from the R package *MuMIn* v1.43.17[61]. The best model was identified (hereafter the quarterly model) and then further applied to the four most annual-rich families (*Asteraceae*, *Brassicaceae*, *Fabaceae*, and *Poaceae*).

Climate Uncertainty

To assess support for our second hypothesis, we investigated the role of IPV and ITV (proxies of climate uncertainty) on annual and annual herb frequencies. We began by testing each variable individually using linear regression and then tested whether their inclusion increased the fit of the quarterly model. Finally, we assessed the increased fit when both IPV and ITV were included in the quarterly model.

Anthropogenic Disturbance

To assess support for our third hypothesis, we measured the impact of the human footprint on annual and annual herb frequencies. We linearly regressed human footprint and annual/annual herbs frequencies and then tested the change in model fit after its inclusion into the quarterly model with IPV and ITV.

Phylogenetic Biases

We applied a phylogenetic generalized least squares regression (pGLS) analysis to account for phylogenetic dependence in the observed patterns. To this end, we devised a continuous response variable for each species by taking the median of the *mean temperature of the warmest quarter* and *precipitation of the warmest quarter* of all their GBIF observations. The explanatory variable was a numeric conversion of each species' life cycle; 1 for annual and 0 for perennial.

The species were matched with those in the GBMB seed plant mega-phylogeny constructed in[62]. The same WFO name resolution process was used on the species in the phylogeny to ensure the same naming scheme. Once matched, we selected the matching herbaceous species resulting in 20,819 species.

The results of the pGLS analysis were compared to the same model without the phylogenetic component (i.e. standard linear regression) to assess the change in coefficient estimates and determine the overall impact of phylogenetic relatedness on our results (see Supplement Note 5).

GBIF Biases

To examine the biases in our dataset with regards to GBIF observational data, we linearly regressed annual and annual herb proportions against the log10 transformed total number of GBIF observations in an ecoregion. Similarly, we conducted a linear regression to assess the relationship between annual and annual herb proportion and the total number of present (5+ observations) GBIF species. Finally, we examined the species in GBIF, but missing from our dataset (see Supplement Note 10 and Extended Data Fig 5).

Spatial Autocorrelation

Following the methods described in[63], spatial eigenvectors for our data were obtained using the R package *adespatial* v0.3.20[64]. We selected the first set of eigenvectors (using those with both positive and negative eigenvalues) that accounted for at least 80% of the variance (39 eigenvectors) and incorporated them into the Yearly-Climate and Quarterly-Climate models. These results were compared to the same models without the eigenvectors included (see Supplement Note 3).

Alternative Regression Models

To ensure that our results are robust to various regression methods, we applied two alternative regression methods. First, we applied a logit transformation to the proportion of annual herbs in each ecoregion followed by linear regression. Second, we used a generalized linear model (Poisson distribution with an offset to represent proportion data) (see Supplement Note 4 and Extended Data Fig 2).

Future Projection

To obtain future projections of the proportion of annual herbs in each ecoregion, future climate estimates in the year 2060 were downloaded from the WorldClim Global Climate Data[59] using the 2041-2060, UKESM1-0-LL[65], ssp585, at ten arc-minutes resolution. The median values for each bioclimatic variable were extracted for each ecoregion using the R package raster v3.4-13[57]. Using the coefficients of a linear regression between the two-most influential climatic parameters found in our study (*mean temperature* and *precipitation during the warmer quarter*, i.e., the Quarterly-Model) and their predicted median value in each ecoregion in 2060, we produced estimates for the proportion of annual herbs in each ecoregion with sufficient data. Year-to-year climate variability and human footprint were not incorporated due to data unavailability at the required resolution and scale. The projected annual herbs proportion in each ecoregion was compared to its current estimate to determine the predicted change in proportion.

Human Footprint Correlations

We examined the relationship between the Human Footprint and various bioclimatic features used in previous analyses. We assessed bioclimatic features 1 & 12 individually and together (the Yearly Model) and bioclimatic features 10 & 12 individually and together (the Quarterly Model) (Supplement Note 12 and Extended Data Fig 6).

## Data and Materials availability

All data and code are available at: https://doi.org/10.6084/m9.figshare.c.6176239

## Methods References

43. Maitner, B. S., *et al*. The bien r package: A tool to access the Botanical Information and Ecology Network (BIEN) database. *Methods Ecol. Evol.* **9,** 373–379 (2018).

44. Tavşanoğlu, Ç., Pausas, J. G. A functional trait database for Mediterranean basin plants. *Sci. Data*. **5,** 180135. figshare https://doi.org/10.6084/m9.figshare.c.3843841.v1 (2018).

45. Parr, C. S., *et al*. The encyclopedia of life v2: providing global access to knowledge about life on earth. *Biodivers. data J.* (2014).

46. Engemann, K., *et al*. A plant growth form dataset for the new world. *Ecology*. **97,** 3243 (2016).

47. WCSP (2021). 'World Checklist of Selected Plant Families. Facilitated by the Royal Botanic Gardens, Kew. Published on the Internet; http://apps.kew.org/wcsp/ Retrieved July 20, 2021.'

48. Kleyer, M., *et al*. The LEDA traitbase: a database of life-history traits of the northwest European flora. *J. Ecol.* **96,** 1266–1274 (2008).

49. Taseski, G. M., *et al*. A global growth-form database for 143,616 vascular plant species. *Ecology*. **53,** 2614 (2019).

50. Kattge, J., *et al*. TRY plant trait database – enhanced coverage and open access. *Glob. Chang. Biol.* **26,** 119–188 (2020).

51. Dauby, G., *et al*. RAINBIO: A mega-database of tropical African vascular plants distributions. *PhytoKeys*. **74,** 1–18 (2016).

52. USDA, NRCS. 2022. The PLANTS Database (http://plants.usda.gov, 05/23/2021). National Plant Data Team, Greensboro, NC USA.

53. Kindt, R. WorldFlora: An R package for exact and fuzzy matching of plant names against the World Flora Online taxonomic backbone data. *Applications in Plant Sciences*, **8,** e11388 (2020).

54. GBIF.org (14 September 2021) GBIF Occurrence Download https://doi.org/10.15468/dl.5d7wa2.

55. Zizka, A., *et al*. CoordinateCleaner: standardized cleaning of occurrence records from biological collection databases. Methods in Ecology and Evolution, 10(5):744-751, doi:10.1111/2041-210X.13152, https://github.com/ropensci/CoordinateCleaner (2019).

56. Olson, D. M., *et al*. Terrestrial ecoregions of the world: a new map of life on Earth: a new global map of terrestrial ecoregions provides an innovative tool for conserving biodiversity. *Bioscience*. **51,** 933–938 (2001).

57. Hijmans, R. J. raster: Geographic data analysis and modeling r package Version 3.4-13. https://CRAN.R-project.org/package=raster (2021).

58. Bivand, R., Keitt, T., Rowlingson, B. rgdal: bindings for the 'geospatial' data abstraction library. R package Version 1.5-27. https://CRAN.R-project.org/package=rgdal (2021).

59. Fick, S. E., Hijmans, R. J. WorldClim 2: new 1-km spatial resolution climate surfaces for global land areas. *Int. J. Climatol.* **37,** 4302–4315 (2017).

60. Venter, O., *et al*. Years of change in the global terrestrial human footprint and implications for biodiversity conservation. *Nat. Commun.*, 12558 (2016).

61. Barton, K. Mu-MIn: Multi-model inference. R Package Version 1.43.17. http://R-Forge.R-project.org/projects/mumin/ (2009).

62. Smith, S. A., Brown, J. W. Constructing a broadly inclusive seed plant phylogeny. *American Journal of Botany*, **105,** 302-314 (2018).

63. Dray, S., *et al*. Community ecology in the age of multivariate multiscale spatial analysis.

*Ecological Monographs*. **82,** 257-275 (2012).

64. Dray, S., *et al*. adespatial: Multivariate multiscale spatial analysis. R package version 0.3-21, https://CRAN.R-project.org/package=adespatial (2023).

65. Sellar, A. A., *et al.* UKESM1: Description and evaluation of the U.K. Earth system model. *Journal of Advances in Modeling Earth Systems*. **11,** 4513-4558 (2019).

**Acknowledgments**
We thank Anna Rice for data processing support and Ron Milo for manuscript critique and feedback. We appreciate the statistical advice and support Micha Mende provided. We are grateful for the insightful comments and critiques offered by the anonymous reviewer, Dr. Neves, and Dr. Salguero-Gomez. We also thank the Edmond J. Safra Center for Bioinformatics at Tel-Aviv University and the ISF for providing funding for this research.

**Funding**
Edmond J. Safra Center for Bioinformatics at Tel-Aviv University
ISF grant no. 672/22 to Niv DeMalach

**Author Contributions**
  Conceptualization: T.P., I.M., N.D.
  Data Curation: T.P.
  Formal Analysis: T.P.
  Funding Acquisition: T.P., I.M., N.D.
  Investigation: T.P.
  Methodology: T.P.
  Project Administration: I.M., N.D.
  Resources: I.M.
  Software: T.P.
  Supervision: I.M., N.D
  Validation: T.P.
  Visualization: T.P.
  Writing – Original Draft Preparation: T.P.
  Writing – Review & Editing: T.P., I.M., N.D.

**Competing Interests**
  Authors declare no competing interests.

**Additional Information**
  Supplementary Information is available for this paper.

  Correspondence and requests for materials should be addressed to Itay Mayrose, itaymay@tauex.tau.ac.il and Niv DeMalach, niv.demalach@mail.huji.ac.il.

Peer review information

Reprints and permissions information is available at www.nature.com/reprints.

# Supplementary Materials for

Revising the global biogeography of plant life cycles

Tyler Poppenwimer[1,2], Itay Mayrose[1*], Niv DeMalach[2†]

*Correspondence to. itaymay@tauex.tau.ac.il
†Correspondence to: Niv.demalach@mail.huji.ac.il

# Supplementary Notes
## Supplement Note 1: Annuals among Herbaceous Species – Results

Here, we present the full results of the analyses presented in the main text for annuals among herbaceous species.

| Yearly Model<br>*Annuals among Herbs* | *Estimate* | *Std. Error* | *P-Value* |
|---|---|---|---|
| (Intercept) | 1.81E-01 | 1.20E-02 | $< 1.0^{-15}$ |
| BioClim1 | 1.12E-02 | 6.41E-04 | $< 1.0^{-15}$ |
| BioClim12 | -7.07E-05 | 1.57E-05 | 8.00E-06 |
| BioClim1 × BioClim12 | -2.13E-06 | 6.80E-07 | 1.81E-03 |
| R-Squared | 0.481 | | |
| DF | 678 | | |
| P-Value | $< 1.0^{-15}$ | | |

| Quarterly Model<br>*Annuals among Herbs* | *Estimate* | *Std. Error* | *P-Value* |
|---|---|---|---|
| (Intercept) | 0.28 | 0.08 | $< 1.0^{-15}$ |
| BioClim10 | 0.01 | 0.00 | $< 1.0^{-15}$ |
| log10(BioClim18 + 1) | -0.13 | 0.03 | $< 1.0^{-15}$ |
| BioClim10 × log10(BioClim18 + 1) | 0.00 | 0.00 | 0.49 |
| R-Squared | 0.547 | | |
| DF | 678 | | |
| P-Value | $< 1.0^{-15}$ | | |

| Temperature Variability<br>*Annuals among Herbs* | *Estimate* | *Std. Error* | *P-Value* |
|---|---|---|---|
| (Intercept) | 2.77E-01 | 1.41E-02 | $< 1.0^{-15}$ |
| T_Var | -8.44E-04 | 2.31E-04 | 2.71E-04 |
| R-Squared | 0.019 | | |
| DF | 679 | | |
| P-Value | 2.71E-04 | | |

| **Precipitation Variability** *Annuals among Herbs* | *Estimate* | *Std. Error* | *P-Value* |
|---|---|---|---|
| (Intercept) | 5.84E-02 | 1.25E-02 | 3.80E-06 |
| P_Var | 1.00E-02 | 6.75E-04 | $< 1.0^{-15}$ |
| R-Squared | 0.244 | | |
| DF | 679 | | |
| P-Value | $< 1.0^{-15}$ | | |

| **Human Footprint** *Annuals among Herbs* | *Estimate* | *Std. Error* | *P-Value* |
|---|---|---|---|
| (Intercept) | 1.75E-01 | 1.11E-02 | $< 1.0^{-15}$ |
| HumanF | 2.69E-03 | 4.76E-04 | 2.43E-08 |
| R-Squared | 0.045 | | |
| DF | 680 | | |
| P-Value | 2.43E-08 | | |

| **All together** *Annuals among Herbs* | *Estimate* | *Std. Error* | *P-Value* |
|---|---|---|---|
| (Intercept) | 2.97E-01 | 9.41E-01 | 7.53E-01 |
| BioClim10 | 4.75E-02 | 4.47E-02 | 2.88E-01 |
| log10(BioClim18 + 1) | -5.35E-02 | 4.07E-01 | 8.95E-01 |
| T_Var | 4.81E-03 | 1.53E-02 | 7.52E-01 |
| P_Var | -8.03E-03 | 4.91E-02 | 8.70E-01 |
| HumanF | 1.17E-02 | 6.04E-02 | 8.47E-01 |
| BioClim10 × log10(BioClim18 + 1) | -2.17E-02 | 1.92E-02 | 2.58E-01 |
| BioClim10 × T_Var | -8.69E-04 | 7.82E-04 | 2.67E-01 |
| log10(BioClim18 + 1) × T_Var | -3.19E-03 | 6.83E-03 | 6.41E-01 |
| BioClim10 × P_Var | -7.60E-04 | 2.15E-03 | 7.24E-01 |
| log10(BioClim18 + 1) × P_Var | 1.90E-03 | 2.26E-02 | 9.33E-01 |
| T_Var × P_Var | -1.37E-04 | 9.25E-04 | 8.82E-01 |
| BioClim10 × HumanF | -1.34E-03 | 2.53E-03 | 5.97E-01 |
| log10(BioClim18 + 1) × HumanF | -5.94E-03 | 2.52E-02 | 8.14E-01 |
| T_Var × HumanF | -4.94E-04 | 1.04E-03 | 6.36E-01 |
| P_Var × HumanF | 1.54E-03 | 3.08E-03 | 6.17E-01 |
| BioClim10 × log10(BioClim18 + 1) × T_Var | 4.35E-04 | 3.45E-04 | 2.08E-01 |
| BioClim10 × log10(BioClim18 + 1) × P_Var | 5.76E-04 | 9.83E-04 | 5.58E-01 |
| BioClim10 × T_Var × P_Var | 2.54E-05 | 3.97E-05 | 5.22E-01 |
| log10(BioClim18 + 1) × T_Var × P_Var | 9.03E-05 | 4.29E-04 | 8.33E-01 |
| BioClim10 × log10(BioClim18 + 1) × HumanF | 7.51E-04 | 1.06E-03 | 4.78E-01 |
| BioClim10 × T_Var × HumanF | 3.78E-05 | 4.53E-05 | 4.04E-01 |
| log10(BioClim18 + 1) × T_Var × HumanF | 2.80E-04 | 4.44E-04 | 5.29E-01 |
| BioClim10 × P_Var × HumanF | -4.20E-05 | 1.23E-04 | 7.32E-01 |
| log10(BioClim18 + 1) × P_Var × HumanF | -5.81E-04 | 1.37E-03 | 6.72E-01 |
| T_Var × P_Var × HumanF | -5.69E-06 | 5.75E-05 | 9.21E-01 |
| BioClim10 × log10(BioClim18 + 1) × T_Var × P_Var | -1.40E-05 | 1.82E-05 | 4.44E-01 |
| BioClim10 × log10(BioClim18 + 1) × T_Var × HumanF | -2.08E-05 | 1.93E-05 | 2.82E-01 |
| BioClim10 × log10(BioClim18 + 1) × P_Var × HumanF | 3.44E-06 | 5.46E-05 | 9.50E-01 |
| BioClim10 | -2.20E-07 | 2.28E-06 | 9.23E-01 |
| $\times$ T_Var $\times$ P_Var $\times$ HumanF | | | |
| log10(BioClim18 + 1) $\times$ T_Var $\times$ P_Var $\times$ HumanF | -2.56E-06 | 2.58E-05 | 9.21E-01 |
| BioClim10 $\times$ log10(BioClim18 + 1) $\times$ T_Var $\times$ P_Var $\times$ HumanF | 4.90E-07 | 1.02E-06 | 6.32E-01 |
| R-Squared | 0.633 | | |
| DF | 649 | | |
| P-Value | $< 1.0^{-15}$ | | |

**Supplement Note 2: <u>Annuals among All Species – Results</u>**

Here, we investigated the proportion of annuals among all species rather than among herbaceous plants (as we did in the main text) (Extended Data Fig 1). Similar to the results presented in the main text, we found that ecoregions with lower precipitation and hotter temperatures (i.e., located in the lower-left coordinate of their biome in Extended Data Fig 1D) possess higher proportions of annuals.

This pattern was corroborated using a linear regression model fitting the annual proportion as a function of *mean yearly temperature* and *total yearly precipitation* ($P < 1.0^{-15}$, $D.F. = 719$, $R^2 = 0.34$) (Extended Data Fig 1E). As in the analysis presented in the main text, using bioclimatic features that account for temporal variation in climate throughout the year produced a better-fitting model. The regression model that incorporated the *mean temperature of the warmest quarter* and the log-transformed *precipitation of the warmest quarter* (Extended Data Fig 1F & G) accounted for 47% of the observed variance ($P < 1.0^{-15}$, $D.F. = 719$) and outperformed the model based on yearly means in terms of information criteria ($\Delta AICc = 160$).

Qualitatively similar relationships between quarterly climate and annual proportions were found in the four most annual-rich families (Asteraceae, Brassicaceae, Fabaceae, and Poaceae). in all families. The explained variation of annual herb proportions ranged from 49% in Brassicaceae to 18% in Fabaceae (in all models $P < 10^{-15}$).

We found that increasing climate unpredictability is associated with a higher proportion of annuals for both precipitation variability ($P < 1.0^{-15}$, $D.F. = 720$, $R^2 = 0.18$) and temperature variability ($P < 1.16^{-5}$, $D.F. = 720$, $R^2 = 0.03$). Incorporating these features into the quarterly model, which included quarterly temperature and precipitation, further improved the model's fit ($P < 1.16^{-5}$, $D.F. = 706$, change in $R^2$ from 0.47 to 0.58, $\Delta AICc = 143$). Finally, the impact of human disturbance was positively correlated, albeit weakly, with a higher proportion of annuals ($P < 0.0103$, $D.F. = 721$, $R^2 = 0.01$). Still, adding this variable to the model with climate unpredictability and quarterly temperature and precipitation further improved the explanatory power of the model ($P < 1.0^{-15}$, $D.F. = 690$, change in $R^2$ from 0.58 to 0.62, $\Delta AICc = 35$).

| **Yearly Model**<br>***Annuals among all Species*** | *Estimate* | *Std. Error* | *P-Value* |
|---|---|---|---|
| (Intercept) | 1.26E-01 | 9.34E-03 | 3.06E-37 |
| BioClim1 | 5.08E-03 | 4.93E-04 | 2.73E-23 |
| BioClim12 | -2.19E-05 | 1.22E-05 | 7.34E-02 |
| BioClim1 × BioClim12 | -2.58E-06 | 5.22E-07 | 9.61E-07 |
| R-Squared | 0.341 | | |
| DF | 719 | | |
| P-Value | 8.69E-65 | | |

| **Quarterly Model** *Annuals among all Species* | *Estimate* | *Std. Error* | *P-Value* |
|---|---|---|---|
| (Intercept) | 2.03E-01 | 5.73E-02 | 4.19E-04 |
| BioClim10 | 9.41E-03 | 2.18E-03 | 1.86E-05 |
| log10(BioClim18 + 1) | -6.56E-02 | 2.55E-02 | 1.04E-02 |
| BioClim10 × log10(BioClim18 + 1) | -2.64E-03 | 9.78E-04 | 7.08E-03 |
| R-Squared | 0.472 | | |
| DF | 719 | | |
| P-Value | 3.2E-99 | | |

| **Temperature Variability** *Annuals among all Species* | *Estimate* | *Std. Error* | *P-Value* |
|---|---|---|---|
| (Intercept) | 8.45E-02 | 9.45E-03 | 3.10E-18 |
| T_Var | 6.90E-04 | 1.56E-04 | 1.16E-05 |
| R-Squared | 0.026 | | |
| DF | 720 | | |
| P-Value | 1.16E-05 | | |

| **Precipitation Variability** *Annuals among all Species* | *Estimate* | *Std. Error* | *P-Value* |
|---|---|---|---|
| (Intercept) | 2.51E-02 | 8.61E-03 | 3.64E-03 |
| P_Var | 5.68E-03 | 4.58E-04 | 3.60E-32 |
| R-Squared | 0.176 | | |
| DF | 720 | | |
| P-Value | 3.6E-32 | | |

| **Human Footprint** *Annuals among all Species* | *Estimate* | *Std. Error* | *P-Value* |
|---|---|---|---|
| (Intercept) | 1.05E-01 | 7.78E-03 | 2.15E-37 |
| HumanF | 8.50E-04 | 3.30E-04 | 1.03E-02 |
| R-Squared | 0.009 | | |
| DF | 721 | | |
| P-Value | 0.010278 | | |

| All together<br>*Annuals among all Species* | *Estimate* | *Std. Error* | *P-Value* |
|---|---|---|---|
| (Intercept) | 5.02E-01 | 6.58E-01 | 4.45E-01 |
| BioClim10 | 1.57E-02 | 3.06E-02 | 6.07E-01 |
| log10(BioClim18 + 1) | -1.64E-01 | 2.84E-01 | 5.63E-01 |
| T_Var | 2.05E-04 | 1.07E-02 | 9.85E-01 |
| P_Var | -2.07E-02 | 3.43E-02 | 5.47E-01 |
| HumanF | 3.42E-02 | 4.17E-02 | 4.13E-01 |
| BioClim10<br>× log10(BioClim18 + 1) | -8.94E-03 | 1.30E-02 | 4.93E-01 |
| BioClim10<br>× T_Var | -3.98E-04 | 5.34E-04 | 4.57E-01 |
| log10(BioClim18 + 1)<br>× T_Var | -6.76E-04 | 4.78E-03 | 8.88E-01 |
| BioClim10<br>× P_Var | -5.25E-06 | 1.48E-03 | 9.97E-01 |
| log10(BioClim18 + 1)<br>× P_Var | 9.45E-03 | 1.58E-02 | 5.50E-01 |
| T_Var<br>× P_Var | 6.42E-05 | 6.51E-04 | 9.21E-01 |
| BioClim10<br>× HumanF | -2.30E-03 | 1.73E-03 | 1.82E-01 |
| log10(BioClim18 + 1)<br>× HumanF | -1.17E-02 | 1.74E-02 | 5.02E-01 |
| T_Var<br>× HumanF | -8.98E-04 | 7.30E-04 | 2.19E-01 |
| P_Var<br>× HumanF | 1.81E-04 | 2.14E-03 | 9.32E-01 |
| BioClim10<br>× log10(BioClim18 + 1)<br>× T_Var | 2.17E-04 | 2.34E-04 | 3.54E-01 |
| BioClim10<br>× log10(BioClim18 + 1)<br>× P_Var | 8.83E-05 | 6.74E-04 | 8.96E-01 |
| BioClim10<br>× T_Var<br>× P_Var | 1.28E-05 | 2.75E-05 | 6.42E-01 |
| log10(BioClim18 + 1)<br>× T_Var<br>× P_Var | -3.37E-05 | 3.01E-04 | 9.11E-01 |
| BioClim10<br>× log10(BioClim18 + 1)<br>× HumanF | 9.19E-04 | 7.21E-04 | 2.03E-01 |
| BioClim10 × T_Var × HumanF | 5.52E-05 | 3.13E-05 | 7.85E-02 |
| log10(BioClim18 + 1) × T_Var × HumanF | 3.62E-04 | 3.12E-04 | 2.46E-01 |
| BioClim10 × P_Var × HumanF | 2.11E-05 | 8.39E-05 | 8.01E-01 |
| log10(BioClim18 + 1) × P_Var × HumanF | -2.00E-04 | 9.42E-04 | 8.32E-01 |
| T_Var × P_Var × HumanF | 1.85E-05 | 4.03E-05 | 6.46E-01 |
| BioClim10 × log10(BioClim18 + 1) × T_Var × P_Var | -6.16E-06 | 1.26E-05 | 6.25E-01 |
| BioClim10 × log10(BioClim18 + 1) × T_Var × HumanF | -2.28E-05 | 1.34E-05 | 8.86E-02 |
| BioClim10 × log10(BioClim18 + 1) × P_Var × HumanF | -1.05E-05 | 3.70E-05 | 7.78E-01 |
| BioClim10 × T_Var × P_Var × HumanF | -1.33E-06 | 1.58E-06 | 4.03E-01 |
| log10(BioClim18 + 1) × T_Var × P_Var × HumanF | -7.31E-06 | 1.81E-05 | 6.86E-01 |
| BioClim10 × log10(BioClim18 + 1) × T_Var × P_Var × HumanF | 6.32E-07 | 7.09E-07 | 3.73E-01 |
| R-Squared | 0.621 | | |
| DF | 690 | | |
| P-Value | 1E-123 | | |

**Supplement Note 3: <u>Spatial Autocorrelation – Results</u>**

Models accounting for spatial autocorrelation incorporated 39 spatial eigenvectors (80.05% of variance explained). The results below are for the proportion of annuals among herbaceous species.

The Yearly Model refers to the model with BioClim1, *mean yearly temperature* and BioClim12, *total yearly precipitation*. The Quarterly Model refers to the model with BioClim10, *mean temperature of the warmest quarter* and BioClim18, *precipitation of the warmest quarter*. Note that BioClim18 is $\log_{10}$ transformed.

The parameter estimates of the yearly and quarterly models show little difference between those models with and without the set of 39 spatial eigenvectors. Similarly, low p-values are associated with each parameter estimate regardless of including the eigenvectors, suggesting negligible differences. These results demonstrate that the yearly and quarterly models are robust to spatial autocorrelation.

| Yearly Model | With Eigenvectors | | | Without Eigenvectors | | |
|---|---|---|---|---|---|---|
| | *Estimate* | *Std. Error* | *P-Value* | *Estimate* | *Std. Error* | *P-Value* |
| Intercept | 0.122 | 0.015 | 2.77E-15 | 0.181 | 1.20E-02 | $< 1.0^{-15}$ |
| BioClim1 | 0.012 | 8.55E-04 | $< 1.0^{-15}$ | 0.011 | 6.41E-04 | $< 1.0^{-15}$ |
| BioClim12 | -6.44E-05 | 1.53E-05 | 2.74E-05 | -7.07E-05 | 1.57E-05 | 8.00E-06 |
| BioClim1 × BioClim12 | -6.56E-07 | 6.91E-07 | 0.343 | -2.13E-06 | 6.80E-07 | 1.81E-03 |
| R-Squared | 0.67 | | | 0.48 | | |
| DF | 639 | | | 678 | | |
| P-Value | $< 1.0^{-15}$ | | | $< 1.0^{-15}$ | | |

| Quarterly Model | With Eigenvectors | | | Without Eigenvectors | | |
|---|---|---|---|---|---|---|
| | *Estimate* | *Std. Error* | *P-Value* | *Estimate* | *Std. Error* | *P-Value* |
| Intercept | 0.119 | 0.077 | 1.20E-01 | 0.276 | 0.077 | 3.60E-04 |
| BioClim1 | 0.015 | 2.94E-03 | 9.10E-07 | 0.014 | 2.94E-03 | 4.79E-06 |
| BioClim18 | -0.065 | 0.033 | 5.12E-02 | -0.128 | 0.034 | 2.04E-04 |
| BioClim1 × BioClim18 | -1.20E-03 | 1.28E-03 | 0.349 | -9.10E-04 | 1.32E-03 | 0.490 |
| R-Squared | 0.67 | | | 0.55 | | |
| DF | 639 | | | 678 | | |
| P-Value | $< 1.0^{-15}$ | | | $< 1.0^{-15}$ | | |

**Supplement Note 4: <u>Alternative Regression Models – Results</u>**

We applied two alternative regression models to the proportion of annual herbs in ecoregions. First, we applied a logit transformation to the proportion of annual herbs (0.01 was added to all annual herb proportions to avoid 0 values) followed by a linear regression (Extended Data Fig 2A & B).

The Yearly Model refers to the model with BioClim1, *mean yearly temperature* and BioClim12, *total yearly precipitation*. The Quarterly Model refers to the model with BioClim10, *mean temperature of the warmest quarter* and BioClim18, *precipitation of the warmest quarter*. Note that BioClim18 is $\log_{10}$ transformed.

| Yearly Model (*Logit Transformed*) | *Estimate* | *Std. Error* | *P-Value* |
|---|---|---|---|
| Intercept | -1.817 | 0.085 | $< 1.0^{-15}$ |
| BioClim1 | 0.080 | 4.52E-03 | $< 1.0^{-15}$ |
| BioClim12 | -3.45E-04 | 1.11E-04 | 1.92E-03 |
| BioClim1 × BioClim12 | -1.93E-05 | 4.80E-06 | 6.55E-05 |
| R-Squared | 0.457 | | |
| DF | 678 | | |
| P-Value | $< 1.0^{-15}$ | | |

| Quarterly Model (*Logit Transformed*) | *Estimate* | *Std. Error* | *P-Value* |
|---|---|---|---|
| Intercept | -0.814 | 0.591 | 0.169 |
| BioClim10 | 0.065 | 0.023 | 3.95E-03 |
| BioClim18 | -0.989 | 2.64E-01 | 1.89E-04 |
| BioClim10 × BioClim18 | 5.89E-03 | 0.010 | 5.61E-01 |
| R-Squared | 0.440 | | |
| DF | 678 | | |
| P-Value | $< 1.0^{-15}$ | | |

Second, we used a generalized linear model with a Poisson distribution and an offset to represent proportion data (Extended Data Fig 2C & D). Note that BioClim18 is $\log_{10}$ transformed.

| Yearly Model (*Poisson GLM*) | *Estimate* | *Std. Error* | *P-Value* |
|---|---|---|---|
| Intercept | -1.839 | 0.015 | $< 1.0^{-15}$ |
| BioClim1 | 0.053 | 8.33E-04 | $< 1.0^{-15}$ |
| BioClim12 | -4.67E-05 | 1.79E-05 | 9.12E-03 |
| BioClim1 × BioClim12 | -2.09E-05 | 8.70E-07 | $< 1.0^{-15}$ |

| Quarterly Model (*Poisson GLM*) | *Estimate* | *Std. Error* | *P-Value* |
|---|---|---|---|
| Intercept | -0.235 | 0.081 | 3.79E-03 |
| BioClim10 | 5.12E-03 | 3.19E-03 | 0.109 |
| BioClim18 | -0.996 | 0.036 | $< 1.0^{-15}$ |
| BioClim10 × BioClim18 | 0.019 | 1.44E-03 | $< 1.0^{-15}$ |

**Supplement Note 5: <u>Phylogenetic Relatedness (pGLS) – Results</u>**

The results below used each species' median bioclimatic values as the basis for the continuous response variable. The annual life cycle was incorporated as a dummy variable (0 - perennial, 1 - annual). BioClim10 is the *mean temperature of the warmest quarter* and BioClim18 is the *precipitation of the warmest quarter*. Note that BioClim18 is $\log_{10}$ transformed.

| **BioClim10** | **Estimate** | **Std. Error** | **statistic** | **P - value** |
|---|---|---|---|---|
| **Intercept** | 23.601618 | 7.021844 | 3.3612 | 0.0007775 |
| **Annual** | 2.872388 | 0.091528 | 31.3826 | < 2.2e-16 |
| **R-Squared** | 0.930 | | | |
| **DF** | 20817 | | | |
| **P-Value** | < 2.2e-16 | | | |

| **$\log_{10}$( BioClim18 +1)** | **Estimate** | **Std. Error** | **statistic** | **P - value** |
|---|---|---|---|---|
| **Intercept** | 2.5471175 | 0.8652659 | 2.9437 | 0.003246 |
| **Annual** | -0.1887782 | 0.0091123 | 20.7168 | < 2.2e-16 |
| **R-Squared** | 0.961 | | | |
| **DF** | 20817 | | | |
| **P-Value** | < 2.2e-16 | | | |

**Supplement Note 6: Database Access Dates**

The following table indicates the dates the respective data was downloaded.

| Database Name | Access | Citation |
|---|---|---|
| BIEN | 01 June 2021 | Maitner, B. S., et al. The bien r package: A tool to access the Botanical Information and Ecology Network (BIEN) database. Methods Ecol. Evol. 9, 373–379 (2018). |
| BROT 2.0 | 23 May 2021 | Tavşanoğlu, Ç., Pausas, J. G. A functional trait database for Mediterranean Basin plants. Sci. Data. 5, 180135. figshare https://doi.org/10.6084/m9.figshare.c.3843841.v1 (2018). |
| EOL | 24 May 2021 | Parr, C. S., et al. The encyclopedia of life v2: providing global access to knowledge about life on earth. Biodivers. data J. (2014). |
| Engemann *et al.*, 2016 | 03 June 2021 | Engemann, K., et al. A plant growth form dataset for the New World. Ecology. 97, 3243 (2016). |
| Kew Gardens WCSP | 20 July 2021 | WCSP (2021). 'World Checklist of Selected Plant Families. Facilitated by the Royal Botanic Gardens, Kew. Published on the Internet; http://apps.kew.org/wcsp/ Retrieved July 20, 2021.' *direct correspondence* |
| LEDA | 01 June 2021 | Kleyer, M., et al. The LEDA Traitbase: a database of life-history traits of the Northwest European flora. J. Ecol. 96, 1266–1274 (2008). |
| Taseski *et al.*, 2019 | 01 June 2021 | Taseski, G. M., et al. A global growth-form database for 143,616 vascular plant species. Ecology. 53, 2614 (2019). |
| TRY | 09 May 2021 | Kattge, J., et al. TRY plant trait database – enhanced coverage and open access. Glob. Chang. Biol. 26, 119–188 (2020). |
| RAINBIO | 06 June 2021 | Dauby, G., et al. RAINBIO: A mega-database of tropical African vascular plants distributions. PhytoKeys. 74, 1–18 (2016). |
| Rice *et al.*, 2019 | 21 June 2021 | Rice, A., et al. The global biogeography of polyploid plants. Nat. Ecol. Evol. 3, 265–273 (2019). |
| USDA | 23 May 2021 | USDA, NRCS. 2022. The PLANTS Database (http://plants.usda.gov, 05/23/2021). National Plant Data Team, Greensboro, NC USA. |

**Supplement Note 7: <u>World Flora Online (WFO) – Filtering</u>**

The following steps were used to filter unreliable or poorly matched species. This process was used as part of the name resolution using the R package *WorldFloraOnline* v1.7[53] as the names database.

1. Species with an *Unchecked* or no taxonomic status were excluded.
2. Species with a Fuzzy Distance (matching distance) greater than two were excluded.
3. Species with a taxonomic rank of *subspecies* or *variety* were united with their respective species, and their taxonomic rank was changed to *species*.
4. Species without a taxonomic rank of *species* were excluded.

**Supplement Note 8: CoordinateCleaner – Filtering**

The following steps were used to discard points with erroneous locations and problematic temporal metadata. This process was applied using *CoordinateCleaner* v2.0-18[55].

1. Data points flagged by the command *clean_coordinates()* using the following capitals
   a. centroids
   b. equal
   c. gbif
   d. institutions
   e. zeros
   f. countries
2. Data points that were flagged by the command *cf_age()* were excluded.

**Supplement Note 9: <u>Gridded System of Polygons – Methods and Results</u>**

In addition to the ecoregion-based system analyses presented in the main text, we repeated the set of analyses using a grid-based system. The grid system was based on a global tessellation of 100km-by-100km square cells using the R package *sf*[66]. First, using a world map provided by the R package *rnaturalearth*[67], we determined the percent land coverage of each cell and excluded those cells with less than 50% land coverage, resulting in 14,595 cells. Next, we mapped the cleaned GBIF observation data points (cleaned as deteilaed in the main text for the ecoregion-based system), into each of these cells using the R packages *raster* v3.4-13[57] and *rgdal* v1.5-27[58]. As with the ecoregion system, we followed the procedures used by [14], whereby species were only considered "present" in a cell if there were five or more observations. Similarly, to ensure all cells contained sufficient data for analysis, each cell was only considered if 10 or more species were present. This procedure produced sufficient data for 5,934 cells (~40.7% of cells) when examining annual species among all species and 5,824 cells (~40.0% of cells) for annual species among only herbaceous species.

We find that cells dominated by annuals among herbaceous species (hereafter just called annuals within this supplemental note) are rare, with only approximately 11.7% exhibiting an annual proportion of 50% or more (Extended Data Fig 4A & B).

First, we projected the proportion of annuals in each grid cell into Whittaker's biomes definitions, as represented using a two-dimensional coordinates system of mean yearly precipitation and temperature. The patterns found using the grid-based system were qualitatively similar but were noisier (probably because ecoregions were designed to minimize variability in environmental conditions within each unit and the smaller number of observations within each grid cell relative to the number of observations within ecoregions). As with ecoregions, cells with lower precipitation and hotter temperatures (i.e., located in the lower-left coordinate of their biome in Extended Data Fig 4C) possess greater percentages of annuals.

This pattern was corroborated using a linear regression model that fitted the proportion of annuals as a function of *mean yearly temperature* and *total yearly precipitation*. Although this gridded system model does not account for as much variation ($P < 10^{-15}$, $D.F. = 5821$, $R^2 = 0.39$) as the ecoregion system model ($R^2 = 0.48$), it nevertheless shows the same trend (Extended Data Fig 4D compared to Fig 2C). Again, bioclimatic features that account for within-year variation in climate produced a model with a better fit. The regression model that incorporated the *mean temperature of the warmest quarter* and the log-transformed *precipitation of the warmest quarter* (Extended Data Fig 4E & F) accounted for 42% of the observed variance ($P < 10^{-15}$, $D.F. = 5821$) and outperformed the annual climate model also in terms of information criteria ($\Delta AICc = 227$).

We also compared the addition of climate unpredictability and human footprint into the models to assess their impact on annual prevalence. We found that increasing climate unpredictability is associated with a higher proportion of annual species ($P < 10^{-15}$, $D.F. = 5821$, $R^2 = 0.19$). Similarly, adding this feature to the model that included quarterly temperature and precipitation further improved the model's fit ($P < 10^{-15}$, $D.F. = 5817$, change in $R^2$ from 0.42 to 0.44, $\Delta AICc = 245$). Lastly, the impact of human disturbance was positively correlated with a higher proportion of annuals ($P < 10^{-15}$, $D.F. = 5821$, $R^2 = 0.16$). Adding this variable to the model with climate unpredictability and quarterly temperature and precipitation further improved the model's explanatory power ($P < 10^{-15}$, $D.F. = 5809$, change in $R^2$ from 0.44 to 0.45, $\Delta$AICc = 100).

66. Edzer, P. Simple features for R: standardized support for spatial vector data. R Package Version 1.0-8. https://cran.r-project.org/web/packages/sf/. (2018).

67. Massicotte, P., South, A. rnaturalearth: world map data from natural Earth. R package version 0.3.0. https://cran.r-project.org/web/packages/rnaturalearth/index.html. (2023).

**Supplement Note 10: <u>GBIF Biases – Results</u>**

We assessed the biases in our dataset with regards to GBIF observational data by first examining the spatial distribution of species missing from our dataset followed by analyses to determine the impact of the number of GBIF species and observations on annual and annual herb proportions.

When examining the proportion of species that were found in GBIF (hereafter "GBIF species") but missing from our dataset in each mapped ecoregion, we found no identifiable regions with disproportionately high missing species (Extended Data Fig 5A). Furthermore, when projecting ecoregions to their respective Whittaker Biomes, we find that no biome missed substantially more species than others before and after the GBIF filtering procedure (Extended Data Fig 5B). Finally, we found that there are very weak correlations between the proportion of GBIF species missing from our dataset and the proportion of annuals among all species ($P = 1.317^{-7}$, $D.F. = 721$, $R^2 = 0.038$) or the proportion of annuals among herbaceous species ($P = 1.182^{-6}$, $D.F. = 680$, $R^2 = 0.034$).

Second, our analyses indicated that there is no correlation between the total number of present (5+ observations) GBIF species and the proportion of annuals among all species ($P = 0.71$, $D.F. = 721$, $R^2 < 0.001$) or the proportion of annuals among herbaceous species ($P = 0.45$, $D.F. = 680$, $R^2 < 0.001$) (Extended Data Fig 5D & F).

Additionally, we find very weak correlations between the ($\log_{10}$ transformed) total number of observations in an ecoregion and the proportion of annuals among all species ($P = 2.101^{-5}$, $D.F. = 721$, $R^2 = 0.025$) or the proportion of annuals among herbaceous species ($P = 0.65$, $D.F. = 680$, $R^2 < 0.001$) (Extended Data Fig 5E & G).

**Supplement Note 11: <u>Biases in Missing Phylogenetic Data – Results</u>**

We explored the biases in our data with regard to the phylogenetic distribution of species missing from our dataset. To this end, we compared the number of species in each family for each non-bryophyte family in the World Flora Online (WFO), which contain 449 families (Extended Data Fig 5H).

- Out of these 449 families:
    - For 288 families, less than 25% of the data is missing.
    - For 393 families, less than 50% of the data is missing.
- The mean percentage of missed species in our dataset is 22.2% per family and the median percentage is 16.2%.

**Supplement Note 12: <u>Human Footprint Correlations</u>**

We found weak correlations between human footprint and the bioclimatic features (individually and in combination) used in the main text (Extended Data Fig 6).

BioClim1 is the *mean yearly temperature* and BioClim12 is the *total yearly precipitation* which are the features in the Yearly Model. BioClim10 is the *mean temperature of the warmest quarter* and BioClim18 is the *precipitation of the warmest quarter* which are the features in the Quarterly Model.

- BioClim1: ($P = 1.0^{-15}$, $D.F. = 721$, $R^2 = 0.15$)
- BioClim12: ($P = 8.8^{-9}$, $D.F. = 721$, $R^2 = 0.04$)
- BioClim1 × BioClim12: ($P = 1.0^{-15}$, $D.F. = 719$, $R^2 = 0.17$)
- BioClim10: ($P = 1.0^{-15}$, $D.F. = 721$, $R^2 = 0.13$)
- Log10(BioClim18+1): ($P = 2.2^{-4}$, $D.F. = 721$, $R^2 = 0.02$)
- BioClim10 × log10(BioClim18+1): ($P = 1.0^{-15}$, $D.F. = 719$, $R^2 = 0.15$)

# Extended Data Items for

# Revising the global biogeography of plant life cycles

Tyler Poppenwimer[1,2], Itay Mayrose[1*], Niv DeMalach[2†]

*Correspondence to. itaymay@tauex.tau.ac.il
†Correspondence to: Niv.demalach@mail.huji.ac.il

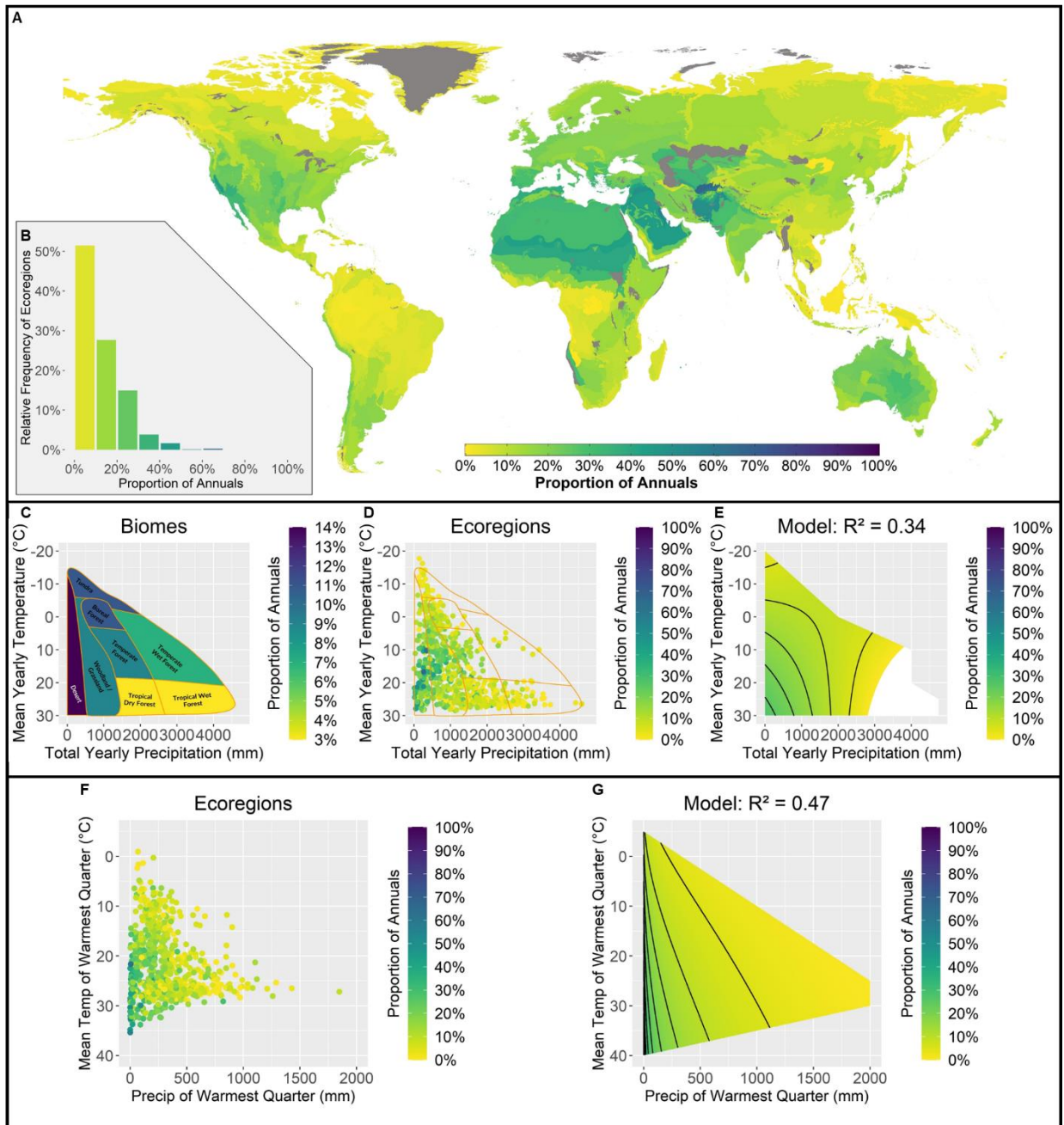

**Extended Data Figure 1 | The global biogeography of the proportion of annual species and the effects of yearly and quarterly climate patterns. A**, the global proportion of annuals in ecoregions with sufficient data; ecoregions with insufficient data (see Methods) are colored grey resulting in 723 colored cells. **B**, The distribution of annual proportions among ecoregions. **C**, The annual proportions in each of Whittaker's Biomes. **D**, A scatterplot of the effects of mean yearly precipitation and temperature (the outline of Whittaker's biomes is marked by orange lines). **e**, Predictions of a regression model of the annual proportions as a function of mean yearly temperature and precipitation (contour lines every 5%). **F**, A scatterplot of the effects of total precipitation and mean temperature of the warmest quarter. **G**, Predictions of a regression model of the annual proportions as a function of total precipitation and mean temperature of the warmest quarter (contour lines every 5%). Note that the scale is different for panel **C.** N=723.

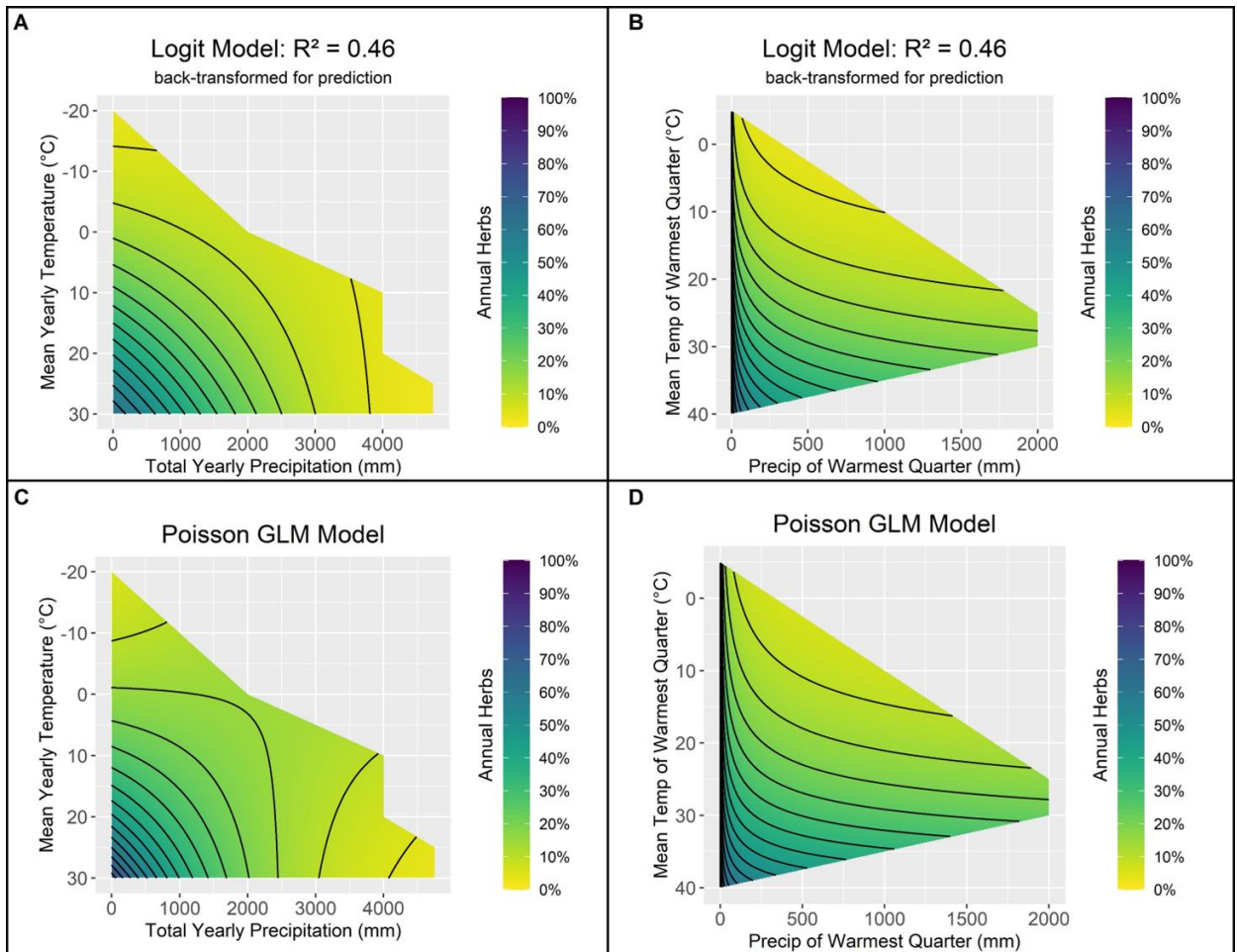

**Extended Data Figure 2 | The predictions of alternative regression methods for the yearly and quarterly models. A and B** depict a logit transformation for the proportion of annual herbs, with **a** showing the yearly model and **B** showing the quarterly model. **C and D** depict the predictions of a Poisson Generalized Linear Model (GLM) with an offset ($\log_{10}$ of the species in an ecoregion) to obtain a rate for the proportion of annual herbs, with **c** showing the yearly model and **D** showing the quarterly model. Contour lines are every 5%

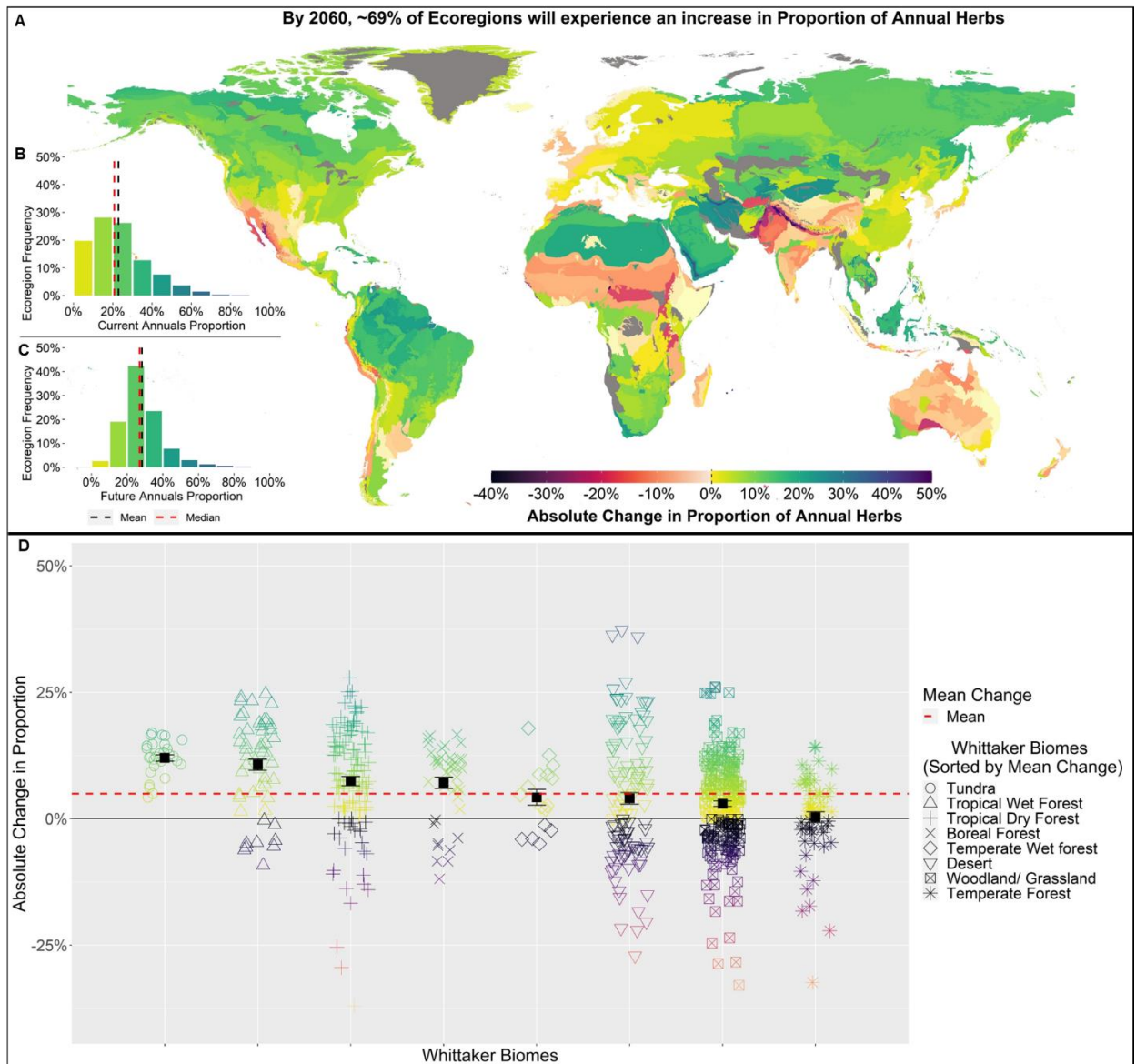

**Extended Data Figure 3 | The predicted change in the proportion of annual herbs in 2100 based on expected climate patterns. A**, The absolute change in the proportion of annual herbs in ecoregions with sufficient data; those with insufficient data are colored grey, resulting in 723 ecoregions. **B**, The current distribution of annual herbs proportions among ecoregions with the mean and median marked by vertical lines. **C**, The predicted future distribution of annual herbs proportions among ecoregions with mean and median demarcated. **D**, The absolute change in the proportion of annual herbs in ecoregions grouped by Whittaker Biome (the color scale is the same as in **A-C**). N=723.

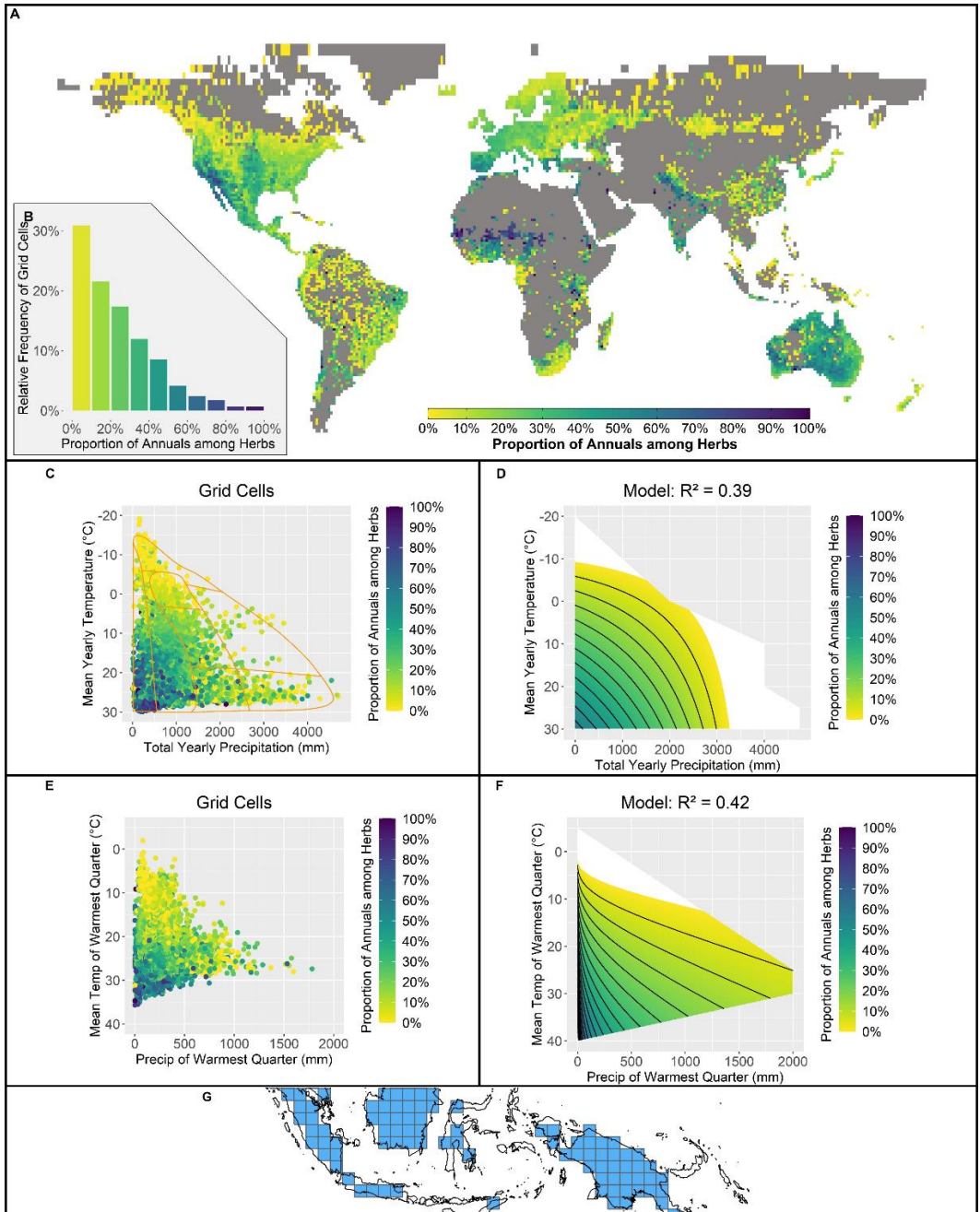

**Extended Data Figure 4 | The global biogeography of the proportion of annual herbs and the effects of yearly and quarterly climate patterns using a gridded system. A**, The global proportion of annuals herbs in cells with sufficient data; cells with insufficient data (see Methods) are colored grey resulting in 5,824 colored cells. **B**, The distribution of annual herbs proportions among grid cells. **C**, A scatterplot of the effects of mean yearly precipitation and temperature (the outline of Whittaker's biomes is marked by orange lines). **D**, Predictions of a regression model of the annual herbs proportion as a function of mean yearly temperature and precipitation (contour lines every 5%). **E**, A scatterplot of the effects of total precipitation and mean temperature of the warmest quarter. **F**, Predictions of a regression model of the annual herbs proportions as a function of total precipitation and mean temperature of the warmest quarter (contour lines every 5%). N=5,824. **G**, A sample of how enlarged grid cells (150km × 150km) omit islands/coastal regions.

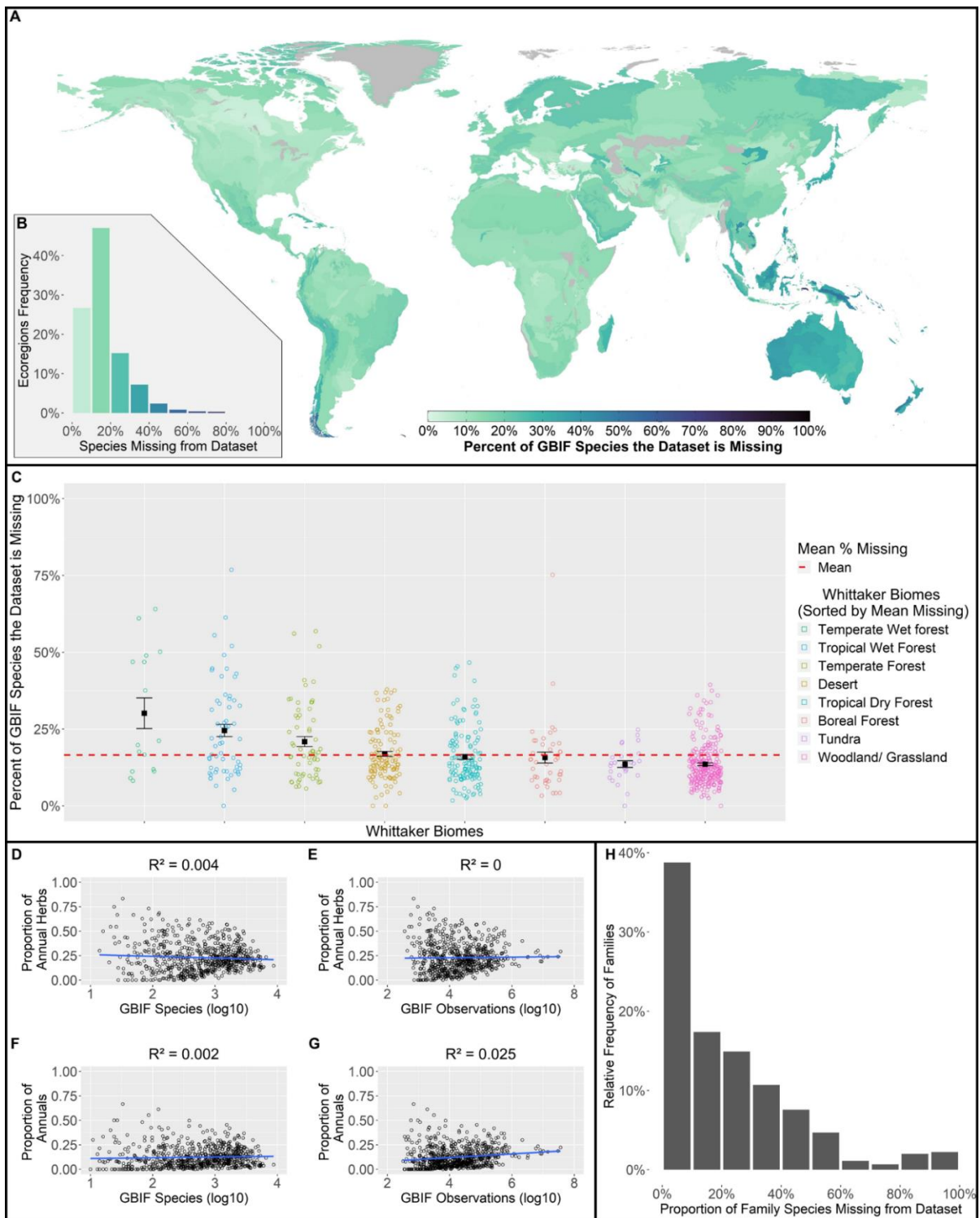

**Extended Data Figure 5 | An exploration of the potential biases in the dataset. A-C** depicts the proportion of present GBIF species the dataset is missing, **A** as a global distribution map, **B** as a distribution of ecoregions, and **c** organized according to Whittaker Biomes. **D-F** shows the correlation of annual herbs (**D and E**) and annuals (**F and G**) with the number of GBIF species and observations. The blue lines depict the best-fit line. **H**, The distribution of the proportion of species in each family the dataset is missing.

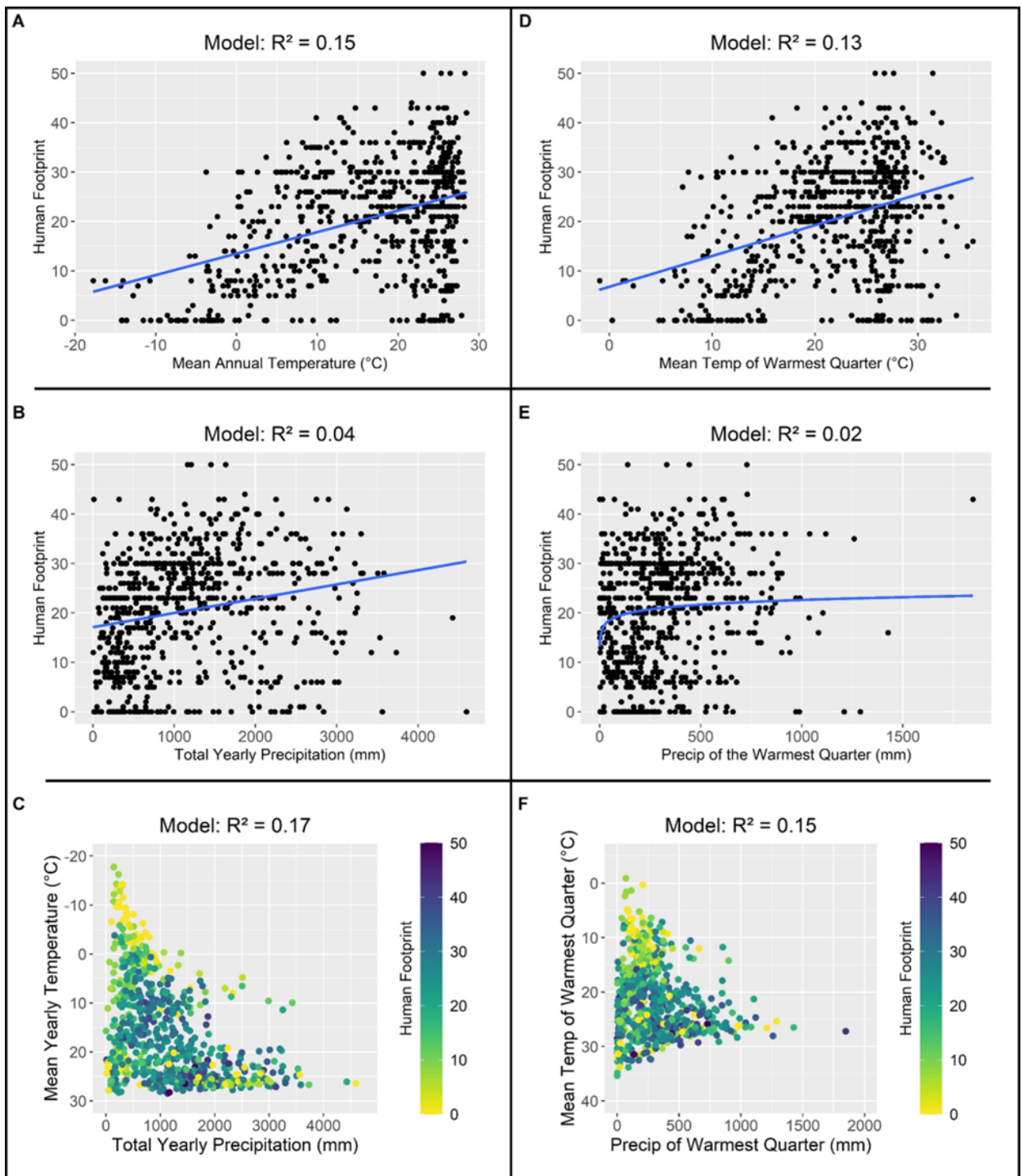

**Extended Data Figure 6 | Scatter plots depicting the correlation between the Human Footprint and various BioClim Features. A-C** show the correlation of the Human Footprint with yearly temperature and precipitation individually **A, B** and together **C**. **D-F** show the correlation of the Human Footprint with quarterly temperature and precipitation individually **D, E** and together **F**. The blue lines depict the best-fit line. Note that the correlation in **E and F** are fitted to the log transformation of the quarterly precipitation.

**Extended Data Table 1 | A comparison of previous estimates, obtained from[4], for the proportion of annuals among all species and among herbs to our revised estimates.** Greyed cells have no initial biome estimate. Alternative previous estimates from[3] are available in Table 1. Note, the biome nomenclature used for the previous estimates differs from ours and so the location of the original study was used to determine our corresponding biome. Additional information can be found in Extended Data Table 4.

| Region | Annuals among All Species | | Annuals among Herbaceous Species | |
|---|---|---|---|---|
|  | Previous Estimate | **Revised Estimate** | Previous Estimate | **Revised Estimate** |
| **Global** | **13%** | **6%** | **29%** | **13%** |
| Desert | 73% or 27% | **14%** | 76%, 66% | **25%** |
| Tundra | 2% | **11%** | 3% | **13%** |
| Woodland / Grassland | 4%, 6%, 14% | **9%** | 11%, 13%, 16% | **19%** |
| Boreal forest |  | **11%** |  | **14%** |
| Tropical Dry Forest | 0% | **3%** | 0% | **14%** |
| Tropical Wet Forest | 10% | **3%** | 59% | **14%** |
| Temperate Forest | 7%, 2% | **9%** | 10%, 3% | **16%** |
| Temperate Wet Forest |  | **7%** |  | **14%** |

**Extended Data Table 2 | The previous estimates obtained from[3, 4], the location of the original study, our corresponding biome based on the original study location, and the calculations used to determine the proportion of annuals among herbs.** Calculations for annual herbs proportions from[3] were determined from values taken from the original studies. Calculations for annual herbs proportions from[4] were determined from values taken from the textbook (original study values were not available).

| Initial Biome Nomenclature | Original Study Location | Corresponding Biome Nomenclature | Annuals among All Species | Annuals among Herbaceous Species |
|---|---|---|---|---|
| **Begon & Townsend (2021)[3]** | | | | |
| Global | | Global | 13% | $\frac{13}{3+27+3+1+13} \approx 27.65\%$ |
| Arctic | Baffin's Island | Tundra | 2% | $\frac{2}{0+51+13+3+2} \approx 2.89\%$ |
| Desert | Death Valley | Desert | 42% | $\frac{42}{0+18+2+5+42} \approx 62.68\%$ |
| Tropical | Seychelles | Wet tropical forest | 16% | $\frac{16}{3+12+3+2+16} \approx 44.44\%$ |
| Temperate | Denmark | Temperate Forest | 18% | $\frac{18}{0+50+11+11+18} = 20\%$ |
| Mediterranean | The Camargue (mouth of the Rhone) | Woodland/ Grassland | 39% | $\frac{39}{0+23+11+39} \approx 53.42\%$ |

Annual herbs calculation: $\frac{Therophyte}{Epiphyte+ Hemicryptophyte+Geophyte+Hydrophyte+Therophyte}$

| Initial Biome Nomenclature | Original Study Location | Corresponding Biome Nomenclature | Annuals among All Species | Annuals among Herbaceous Species |
|---|---|---|---|---|
| **Gurevitch et al. (2021)[4]** | | | | |
| Global | | Global | 13% | $\frac{13}{26+6+13} \approx 28.88\%$ |
| Tropical Rainforest | Queensland | Tropical Dry Forest | 0% | 0% |
| Subtropical forest | Matheran, India | Tropical Wet Forest | 10% | $\frac{10}{2+5+10} \approx 58.82\%$ |
| Warm-temperate Forest | Mediterranean live-oak forest, 0-500 m | Woodland /Grassland | 4% | $\frac{4}{24+9+4} \approx 10.81\%$ |
| Cold-temperate forest | Central Siskiyou Mtns | Temperate Forest | 7% | $\frac{7}{54+12+7} \approx 9.58\%$ |
| Tundra | Spitzbergen | Tundra | 2% | $\frac{2}{60+15+2} \approx 2.59\%$ |
| Mid-temperate mesophytic forest | Siskiyou Mtns South Fork and Beaver Creek | Temperate Forest | 2% | $\frac{2}{33+23+2} \approx 3.44\%$ |
| Oak Woodland | Santa Catalina Mountains, AZ | Woodland /Grassland | 6% | $\frac{6}{36+5+6} \approx 12.76\%$ |
| Dry Grassland | Pamir Mts. Steppe | Woodland /Grassland | 14% | $\frac{14}{63+10+14} \approx 16.09\%$ |
| Semi-desert | Oudjda semi-desert | Desert | 27% | $\frac{27}{14+0+27} \approx 65.85\%$ |
| Desert | Oudjda desert | Desert | 73% | $\frac{73}{17+6+73} \approx 76.04\%$ |

Annual herbs calculation: $\frac{Therophyte}{Hemicryptophyte+Cryptophyte+ Therophyte}$

**Extended Data Table 3 | The original studies of global and biome-level annuals proportion estimates found in[6, 7] their location and sample size.** Note that the origin for the global annual frequency estimate is the same as in Extended Data Table 4.

| Biome | Proportion | Original Study | Location | Species |
|---|---|---|---|---|
| Global | 13% | Raunkiær, C. (1918). Über das biologische Normalspektrum. Andr. Fred. Høst & søn, Bianco Lunos bogtrykkeri. | Worldwide | 400 |
| Tropical | 16% | Raunkiær, C. (1918). Über das biologische Normalspektrum. Andr. Fred. Høst & søn, Bianco Lunos bogtrykkeri. | Seychelles | 258 |
| Desert | 42% | Raunkiær, C. (1918). Über das biologische Normalspektrum. Andr. Fred. Høst & søn, Bianco Lunos bogtrykkeri. | Death Valley | 294 |
| Mediterranean | 39% | Raunkiær, C. (1918). Über das biologische Normalspektrum. Andr. Fred. Høst & søn, Bianco Lunos bogtrykkeri. | The Camargue (mouth of the Rhone) | 233 |
| Temperate | 18% | Raunkiær, C. (1918). Über das biologische Normalspektrum. Andr. Fred. Høst & søn, Bianco Lunos bogtrykkeri. | Denmark | 1084 |
| Arctic | 2% | Raunkiær, C. (1918). Über das biologische Normalspektrum. Andr. Fred. Høst & søn, Bianco Lunos bogtrykkeri. | Baffin's Land | 129 |

**Extended Data Table 4 | The original studies of global and biome-level annuals proportion estimates found in[5, 8], their location and sample size.** Note that the origin for the global annual frequency estimate is the same as in Extended Data Table 3.

| Biome | Proportion | Original Study | Location | Species |
|---|---|---|---|---|
| Global | 13% | Raunkiær, C. (1918). Über das biologische Normalspektrum. Andr. Fred. Høst & søn, Bianco Lunos bogtrykkeri. | Worldwide | 400 |
| Tropical Rainforest | 0% | Cromer, D. A. N., & Pryor, L. D. (1942). A contribution to rain-forest ecology. In Proceedings of the Linnean Society (Vol. 67, pp. 249-268). | Queensland, Australia | 141 |
| Subtropical Forest | 10% | Bharucha, F. R., & Ferreira, D. B. (1941). The biological spectrum of the Madras flora. Journal of University of Bombay, 9, 93-100. | Matheran, India | 361 |
| Warm-temperate Forest | 4% | Braun-Blanquet, J., (1936). La Chênaie d'Yeuse méditerranéenne (Quercion ilicis): monographie phytosociologique. Sta. Internatl. Dr Géobot Méditer. Et Alp. Montpellier Commun. **45**: 1-147 | Mediterranean live-oak forest, 0-500 m | |
| Cold-temperate Forest | 7% | Whittaker, R. H. (1960). Vegetation of the Siskiyou Mountains, Oregon and California. Ecological Monographs, 30(3), 279–338. | Central Siskiyou Mtns., by elevation belts on diorite: 1920-2140 m | 72 |
| Tundra | 2% | Raunkiær, C. (1918). Über das biologische Normalspektrum. Andr. Fred. Høst & søn, Bianco Lunos bogtrykkeri. | Spitzbergen | 110 |
| Mid-temperate Forest | 2% | Whittaker, R. H. (1960). Vegetation of the Siskiyou Mountains, Oregon and California. Ecological Monographs, 30(3), 279–338. | Mixed Evergreen Forest in the West (South Fork and Beaver Creek) | 160 |
| Oak Woodland | 6% | Whittaker, R. H., & Niering, W. A. (1965). Vegetation of the Santa Catalina Mountains, Arizona: a gradient analysis of the south slope. Ecology, 46(4), 429-452. | Santa Catalina Mountains, Arizona | 100 |
| Dry Grassland | 14% | Paulsen, O. (1915). Some remarks on the desert vegetation of America. The Plant World, 18(6), 155-161. | Pamir Mts. Steppe | 514 |
| Semi-desert | 27% | Braun-Blanquet, J., & Maire, R. C. J. E. (1924). Etudes sur la végétation et la flore marocaines. | Oudjda semi-desert | 32 |
| Desert | 73% | Braun-Blanquet, J., & Maire, R. C. J. E. (1924). Etudes sur la végétation et la flore marocaines. | Oudjda desert | 49 |